# Impacts of EPA's Finalized Power Plant Greenhouse Gas Standards

**Authors:** John Bistline[1]*, Aaron Bergman[2], Geoffrey Blanford[1], Maxwell Brown[3], Dallas Burtraw[2], Maya Domeshek[2], Allen Fawcett[4], Anne Hamilton[5], Gokul Iyer[4], Jesse Jenkins[6], Ben King[7], Hannah Kolus[7], Amanda Levin[8], Qian Luo[6], Kevin Rennert[2], Molly Robertson[2], Nicholas Roy[2], Ethan Russell[2], Daniel Shawhan[2], Daniel Steinberg[5], Anna van Brummen[7], Grace Van Horn[9], Aranya Venkatesh[1], John Weyant[10], Ryan Wiser[11], Alicia Zhao[4]

[1] EPRI; Palo Alto, CA, USA.
[2] Resources for the Future; Washington DC, USA.
[3] Colorado School of Mines; Golden, CO, USA.
[4] Center for Global Sustainability, University of Maryland; College Park, MD, USA.
[5] National Renewable Energy Laboratory; Golden, CO, USA.
[6] Princeton University; Princeton, NJ, USA.
[7] Rhodium Group; Oakland, CA, USA.
[8] Natural Resources Defense Council; Washington DC, USA.
[9] Center for Applied Environmental Law and Policy; Washington DC, USA.
[10] Stanford University; Stanford, CA, USA.
[11] Lawrence Berkeley National Laboratory; Berkeley, CA, USA.

*Corresponding author. Email: jbistline@epri.com.

Introduction

The Inflation Reduction Act (IRA) subsidizes the deployment of clean electricity, hydrogen production, and carbon capture and storage (CCS), which could enable additional actions by other federal, state, and local policymakers to reduce emissions. Power plant rules finalized by the Environmental Protection Agency (EPA) in 2024 are one such example of complementary policies. The rules establish emissions intensity standards, not technology mandates, meaning power plant owners can choose from a range of technologies and control options provided that emissions standards are met. This flexibility makes electricity systems modeling important to understand the potential effects of these regulations. We report below a multi-model analysis of the EPA power plant rules that can provide timely information, including for other countries and states, on emissions impacts, policy design for electricity decarbonization, power sector investments and retirements, cost impacts, and load growth. We also discuss related technical, political, and legal uncertainties.

The rules for new gas and existing coal power plants are pursuant to Section 111 of the Clean Air Act, following earlier efforts to regulate greenhouse gas emissions from existing power plants like the Clean Power Plan and Affordable Clean Energy rules, which faced legal challenges or were repealed after changes in the administration (*1*). The rules require power plants to meet emissions thresholds that vary by the plants' retirement dates and operational characteristics. The emission rate limits are based on the "best system of emission reduction" (BSER), which is CCS with 90% capture for existing coal-fired plants operating past 2038 and 90% CCS for new gas-fired plants in 2032 if operating with over 40% utilization (generating at least 40% of their annual maximum capacity) (see supplementary materials (SM)). While the emissions limits are based on specific technological assumptions, power plants can meet or exceed these limits on their average $CO_2$ emissions per unit of electricity generated using a range of options, which could include CCS, cofiring with lower-emitting fuels such as natural gas at coal plants, or efficiency improvements. EPA also created additional subcategories with different thresholds, which means that not all plants are subject to CCS-based standards.

The rules intersect with other power sector trends—growing electricity use from data centers and electrified services, targets to reach net-zero emissions, political uncertainty about federal climate legislation, and grid transitions toward lower-emitting resources. Although coal use has been declining since 2011 (Fig. S5), coal represented nearly half of U.S. power sector $CO_2$ in 2023 despite being only 16% of generation.

Our multi-model analysis delivers timely information on:

- **Emissions impacts of EPA's power plant rules:** The international community and U.S. government are assessing progress toward Paris Agreement goals (*2*) and how much work is left for other federal, state, and company actions and for other sectors after accounting for these rules. Power sector emissions also affect the timing of IRA tax credit expirations (*3-4*).

- **Policy design for electricity decarbonization:** The U.S. electric sector is the second highest greenhouse gas emitting sector in the world's second highest emitting country (*5*). Insights about the cost-effectiveness of policy and technology strategies may be relevant for other countries and subnational jurisdictions, since committed emissions from existing power plants may jeopardize global climate targets without early retirements (*6*) and due to the central roles of decarbonizing electricity and electrifying end uses for net-zero efforts (*7*).

- **Power sector investments and retirements:** This information is valuable to states as they draft plans to comply with EPA's standards for existing coal plants (which are due in May 2026), technology developers and electric company planners evaluating responses, system operators considering reliability implications, and local governments understanding impacts of plant closures on jobs and tax revenue.

- **Cost impacts:** Analysis on policy compliance costs and electricity prices can inform the public about potential impacts of the rules and legal challenges.

- **Load growth:** Utilities, policymakers, and the public are looking to understand system implications of rapid electricity demand growth due to data centers, industrial facilities, and electrification of vehicles and other end-uses (*8*) and how these effects could change with emissions regulations.

Modeling the EPA Rules

This analysis uses nine models of the electric sector and energy systems to understand potential impacts of EPA's finalized power plant rules. By aligning key assumptions and running harmonized scenarios, model comparisons like ours can identify common findings about impacts and quantify levels of disagreement across participating models. We also compare EPA's Regulatory Impact Analysis for the rules (*9*) with our results, thereby providing a richer description and range of possible impacts than a single model can provide.

To evaluate effects on emissions and power sector outcomes, scenarios with EPA's standards for new and existing power plants are compared to reference scenarios without these rules (the article uses "the rules" as shorthand for EPA's finalized new and existing standards) (see SM for detailed descriptions of models and study assumptions). All scenarios include current policies as of early 2024 (Table S1), including major IRA provisions, and harmonized assumptions about technology costs, fuel prices, and financing. Models include a greater range of mitigation options for coal- and gas-fired power plants than earlier studies, including CCS retrofits and cofiring with lower-emitting fuels (Table S2), both to capture compliance pathways in the rules and to understand deployment of IRA-supported resources. Scenarios with and without the rules are run with higher electricity demand levels to understand how additional growth from data centers, manufacturing, and electrification may alter these outlooks. A final sensitivity analysis examines potential impacts of adding standards for existing natural gas-fired power plants, which are not included in the finalized rules.

Emissions Implications

Model results suggest that the finalized rules accelerate emissions reductions in the power sector. The range of projected $CO_2$ emissions is 73-86% below 2005 levels by 2040, compared with 60-83% in the reference without the rules (Fig. 1). The rules narrow the range of potential $CO_2$ emissions and hence can be viewed as backstops against higher emissions outcomes under futures with improved coal plant economics, which could occur with higher demand, slower renewables deployment from interconnection and permitting delays, or higher natural gas prices. The rules cut $CO_2$ by 68-390 Mt-$CO_2$ annually in 2040 compared to current policies (Fig. S4), which are greater reductions than EPA's Regulatory Impact Analysis suggests (54 Mt-$CO_2$).

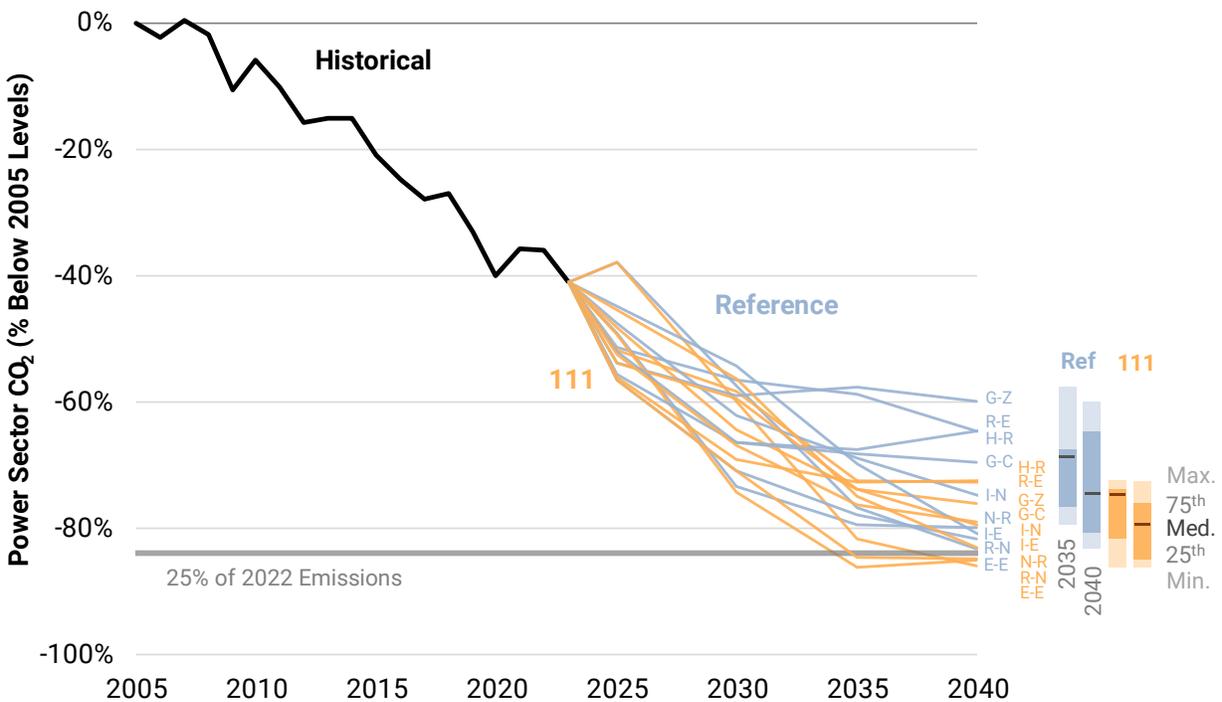

**Fig. 1. Cross-model comparison of U.S. power sector emissions reductions over time with and without EPA's power plant rules.** Ranges on the right show minimum, 25th percentile, median, 75th percentile, and maximum values across models with current policies only ("Reference") and with EPA rules ("111"). Inflation Reduction Act tax credits begin to expire in 2032 or after power sector $CO_2$ reaches 25% of 2022 levels, whichever is later. Details on models and study assumptions are provided in Materials and Methods S1 and S3.

Due to the timelines associated with the rules, which primarily are in the 2030s (SM S2), the rules make limited contributions toward reaching the 2030 U.S. economy-wide emissions target (*2*). Even with the rules, power sector emissions also fall short of a net-zero $CO_2$ goal (*2*), though the rules narrow the implementation gap (Fig. 1). Technology-neutral power sector credits under IRA begin to expire when electricity $CO_2$ emissions reach 25% of 2022 levels. The rules may move forward this date, but seven of the nine models do not cross this point by 2035 and six still fail to do so by 2040.

The rules also accelerate reductions of conventional air pollutants with 88-98% reduction in $SO_2$ by 2035 compared to 2015 (70-88% in the reference) and 84-94% in $NO_x$ emissions (74-90% in the reference) (Fig. S6). These co-pollutant reductions can bring near-term air quality benefits and improve public health, including in environmental justice communities (*10*), which have been disproportionately impacted by pollution.

Effects on Power Sector Capacity and Generation

The rules could have the largest impacts on reducing installed coal capacity and generation (Fig. S10 and Fig. 2, respectively), which historically have been the largest source of power sector $CO_2$ and conventional air pollutants. Coal capacity declined steadily over the past 15 years (Fig. S7), which reflects economic pressures in many markets and retirement announcements linked to company emissions targets. Models indicate that the rules could accelerate coal retirements relative to historical levels and reference trends (Fig. S10), albeit with differences across models in this extent. Compared to EPA's Regulatory Impact Analysis, models in this analysis have fewer coal retirements in the reference scenario by 2040 and lower CCS-equipped coal with the rules (Fig. S20). Coal capacity is replaced by a portfolio that varies by model: Retirements are offset by dispatchable capacity that can adjust output to meet demand (Fig. S9), which is largely new gas-fired units with some energy storage, CCS, and retained nuclear capacity. Gas capacity increases relative to reference levels for many models, even with new source standards, though magnitudes are small in comparison with solar and wind additions (Fig. S10).

Although the standards are based on the application of CCS, the analysis finds limited CCS deployment by 2035 for new gas or existing coal (Fig. S10). Although IRA's credits of up to $85/t-$CO_2$ improve the economics of CCS, most coal capacity retires instead of retrofitting with CCS. CCS-equipped generation is a small part of total generation by 2040 (0.7-3.0%, Fig. 2) and similar to the reference (Fig. S8), suggesting that projects are driven more by IRA incentives and state policy than by EPA rules. These scenarios illustrate the possibility that compliance with the rules could be achieved without incremental CCS.

The rules are more likely to reduce coal than to curb new gas capacity, since new gas-fired units operating at less than 40% annual utilization do not face stringent emissions standards under the rules (Fig. S8). Although new gas dominates capacity changes (Fig. S9), existing natural gas combined cycle (NGCC) plants—which are not regulated under the finalized rules—are the largest substitute for displaced coal generation, in addition to increased renewables, existing nuclear, and CCS. NGCC plants are the most common gas-fired generation resources in the U.S. and use both gas and steam turbines to improve efficiency. The rules increase the utilization of existing NGCC and decrease generation from units covered by regulations such as new gas-fired capacity (Fig. S11). These offsetting effects of different parts of the rules lead to slight increases in gas generation shares for most models under the rules. However, models generally suggest a smaller role for gas in both the reference and 111 scenarios (Fig. 2) compared to its generation share of 42% in 2023 (14-31% in the reference and 15-31% with the rules in 2035). Similarly, average capacity factors of gas-fired capacity are expected to decline relative to today (Fig. S12), and this declining utilization means that many new gas plants opt to run with reduced capacity factors rather than to install CCS. These gas increases are approximately offset by coal generation decreases, which makes changes to overall fossil generation shares under the rules relatively small (Fig. S13).

These capacity changes may have implications for resource adequacy and reliability. Although regional analysis is needed to evaluate such impacts in detail (*11*), the analysis suggests that retiring coal is largely replaced by dispatchable capacity, which means that the direction and magnitude of reliability metrics depend on the relative outage rates of existing coal vis-à-vis natural gas, location of retirements and additions, timing of replacement capacity, and the ability of energy storage and renewables to contribute to system reliability. In addition, the rules contain features to aid reliability during periods of system stress, including options for units to respond to declared system emergencies and to stay online for documented reliability needs, the extension of compliance timelines due to implementation delays, and additional compliance flexibilities.

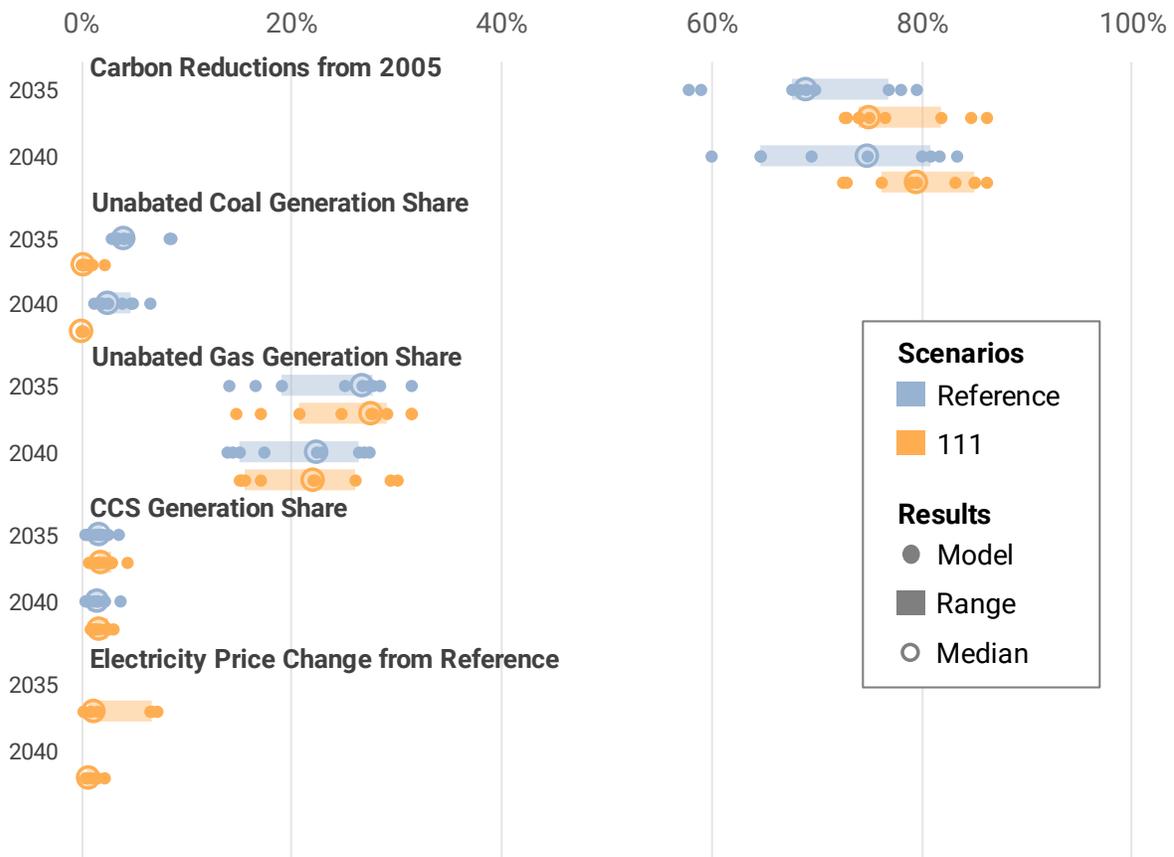

**Fig. 2. Summary of key indicators for 111 and reference scenarios across models.** From the top, indicators are electric sector $CO_2$ reductions (% from 2005 levels), generation share from coal without CCS, generation share from natural gas without CCS, CCS-equipped generation share, and wholesale electricity price changes relative to the reference scenario. For each metric and scenario, values are shown for individual models (dots), median (open circles), and interquartile range across models (bars).

These comparisons highlight a few potential unintended consequences of the rules' design:

- The models project that the rules may cause reductions in generation from covered units (i.e., existing coal and new gas) and increases from uncovered units (i.e., existing gas and renewables). Despite this rebound from existing gas, the overall effect of the rules is to lower $CO_2$ emissions (Fig. S5). The increase in generation from existing gas is larger for many models than EPA's modeling (Fig. S9), which partially reflects more coal retirements in EPA's reference case.

- Some have raised concerns about whether the 40% capacity factor threshold for new gas could lead to more installed capacity that reaches this limit. The analysis shows that new NGCC additions increase under the rules (Fig. S8), though this increase is also caused by retiring coal units from the existing source standards.

Cost Impacts

Model results indicate that the rules may be met with relatively small costs, even before accounting for climate, public health, and other societal benefits. Bulk power system costs increase 0.5-3.7% with the rules through 2050 relative to the reference due primarily to higher investment costs for gas capacity that replaces retiring coal, though there is cross-model variation in the composition of power system expenditures (Fig. S14). Wholesale electricity price changes are similarly low with the rules and decrease over time to less than 2.2% above reference levels in 2040 across all models (Fig. 2). Note that, even if aggregate national costs are low, regional impacts could be larger.

These low system costs mean that abatement costs of these rules are much lower than many recent estimates of the social cost of $CO_2$, cost of controls for other EPA rules, and "best system of emission reduction" costs (Fig. S15). Average abatement costs range from $6-44/t-$CO_2$. These low abatement costs reflect flexibilities in the rules and coal retirements being among the lowest-cost mitigation opportunities due to their ease of substitution (independent of the EPA rules), which aligns with the broader decarbonization literature *(7)*. The consistency of this result is notable given the differences in compliance pathways across models.

Load Growth Impacts

Policymakers and planners expect future electricity demand growth to exceed recent historical levels, though there is uncertainty about contributions from data centers (in part from artificial intelligence applications), domestic manufacturing, IRA-subsidized electrolytic hydrogen production, and end-use electrification (*8*). These trends are reflected in the earlier scenarios, which have 1.3-2.3% annual growth rates in total electricity demand through 2035 (compared to an average 0.2% rate in the 2010s), but we also analyze scenarios with even higher load growth rates (1.8-2.8%) to understand implications for 111 compliance and emissions.

Emissions and generation responses to higher electricity demand vary over time and by model (Fig. S17). Gas-fired resources are generally more responsive to new electricity demand in the near term, but the $CO_2$ intensity of power generation declines over time and is less than the current grid mix for nearly all models and scenarios, especially with the rules. Overall, zero-emitting resources make up 55-83% of increased generation by 2040 under the rules compared with 36-81% by 2035. Incremental $CO_2$ reductions from the rules are similar under reference and higher demand assumptions (2-16 percentage points versus 3-16, respectively, in 2040). $CO_2$ impacts from load growth are generally lower with the rules than without them, though some models show little impact while others show large declines for cases where coal generation increases with higher demand in the reference scenario, which is mitigated by the rules. This finding reinforces the role of these rules as a potential risk reduction tool to prevent emissions rebounds if load growth or other factors increase the competitiveness of coal generation. Note that emissions could be even lower with company procurement goals, greater flexibility of end-use loads, and displaced production outside of the U.S. for manufacturing, which are not included here.

Effects of Future Policies

Although the finalized rules do not cover existing natural gas-fired combustion turbines, EPA could, in the future, undertake rulemakings to address carbon emissions and other pollutants from the existing gas fleet (*1*). To understand how these regulations could alter power sector emissions and other outcomes, we conduct an additional scenario that adopts similar timelines and subcategories for existing gas units as the finalized new source standards for gas. The largest modeled impacts of extending standards to existing gas units is to increase generation from new NGCC with and without CCS, while generation is lower for existing NGCC. However, these changes are relatively small compared to other system changes, which means that incremental $CO_2$ reductions from the modeled existing gas standards are lower than impacts of the finalized rules (Fig. S18). One driver of the more limited emissions response is that a substantial share of existing gas plants opt to operate at capacity factors of 40% or lower rather than retrofitting with CCS or other measures to more substantially reduce their emissions rates, given that gas utilization already declines before accounting for the rules (Fig. S12).

Future policy uncertainty at federal and state levels also may impact the rules. Many state policies and company targets look to reduce emissions to net-zero levels by midcentury across the economy. Since rapid declines in coal consumption are features of such pathways (*7*), these actions could lower the rules' impacts, though many states that currently have net-zero policies and stringent clean electricity standards are not ones with large shares of existing coal capacity.

Additional Uncertainties

In addition to unknowns about load growth and future policies, several uncertainties could alter EPA's power plant rules:

- U.S. presidential and congressional elections and the change in administration may influence whether the EPA power plant regulations remain in place, are revised, or are augmented. There are also unknowns about how congressional elections may shape IRA's tax credits and further legislative climate policy, which may affect technology deployment and compliance under the rules.

- There have already been court challenges to the rules, and although motions to stay the rules have thus far been denied, all parties agreed to an expedited briefing on the merits before the U.S. Court of Appeals for the District of Columbia Circuit that was held on December 6, 2024. As litigation of these rules continues, there is still potential for the courts to stay the rules or to require them to be rewritten in response to legal challenges.

- Ultimately, states have flexibility in developing their state plans to comply with existing source standards for coal plants. States can implement trading, averaging, and several types of flexibility and invoke the remaining useful life and other factors (RULOF) to apply less stringent standards of performance. State plans currently are due to EPA in May 2026.

- Uncertainties around the siting, permitting, and interconnection of power sector resources raise questions about the rate of future change (*12*). The pace of scaling and associated infrastructure buildouts are especially pressing for wind and solar, which are the technologies with the highest deployment rates across all models in our analysis (Fig. S10). These issues are also uncertain for emerging technologies like CCS and hydrogen, which are expected to play key roles in reaching economy-wide net-zero emissions (*13-14*) but require supporting energy economies for potential transport and storage. States can include a mechanism that allows extensions up to one year for unanticipated delays beyond the owner or operator's control such as permitting or supply chain challenges.

**Acknowledgments:** The views and opinions expressed in this paper are those of the authors alone and do not necessarily represent those of their respective institutions, the U.S. Department of Energy (DOE), the U.S. Government, or other funding agencies, and no official endorsement should be inferred. The authors using E4ST thank Gurobi for the use of its solver and Energy Visuals Inc. for the use of its transmission data. A.F. and G.I. are also affiliated with Pacific Northwest National Laboratory, which did not provide specific support for this paper. The AAAS recognizes the U.S. Government's non-exclusive rights to use the Work for non-commercial, governmental purposes where such rights are established in the contract.

**Funding:** J.J. and Q.L. were funded by the Princeton Zero-Carbon Technology Consortium, supported by unrestricted gifts from GE, Google, and Breakthrough Energy, and the Princeton Carbon Mitigation Initiative, supported by a gift from bp. A.B., D.B., M.D., K.R., M.R., N.R., E.R., and D.S. were funded by RFF's Electric Power Program. M.B., A.H., and D.S. were funded by the DOE Office of Policy through the National Renewable Energy Laboratory, operated by Alliance for Sustainable Energy, LLC, for the DOE under Contract no. DE-AC36-08GO28308. R.W. was funded by the DOE Office of Policy under Lawrence Berkeley National Laboratory Contract no. DE-AC02-05CH11231. Data are available at https://doi.org/10.5281/zenodo.14322914.

**Author contributions:**

Conceptualization: All authors

Methodology: All authors

Investigation: All authors

Visualization: JB

Writing – original draft: All authors

Writing – review & editing: All authors

**Competing interests:** J.J. is part owner of DeSolve, LLC, which provides techno-economic analysis and decision support for clean energy technology ventures and investors. A list of clients can be found at https://www.linkedin.com/in/jessedjenkins. He serves on the advisory boards of Eavor Technologies Inc., a closed-loop geothermal technology company, Rondo Energy, a provider of high-temperature thermal energy storage and industrial decarbonization solutions, and Dig Energy, a developer of low-cost drilling solutions for ground-source heat pumps and has an equity interest in each company. He also serves as a technical advisor to MUUS Climate Partners and Energy Impact Partners, both investors in early-stage climate technology companies. The other authors declare no competing interests.

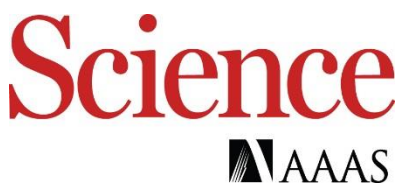

# Supplementary Materials for

### Impacts of EPA's Finalized Power Plant Greenhouse Gas Standards

John Bistline[1]*, Aaron Bergman[2], Geoffrey Blanford[1], Maxwell Brown[3], Dallas Burtraw[2], Maya Domeshek[2], Allen Fawcett[4], Anne Hamilton[5], Gokul Iyer[4], Jesse Jenkins[5], Ben King[7], Hannah Kolus[7], Amanda Levin[8], Qian Luo[6], Kevin Rennert[2], Molly Robertson[2], Nicholas Roy[2], Ethan Russell[2], Daniel Shawhan[2], Daniel Steinberg[5], Anna van Brummen[7], Grace Van Horn[9], Aranya Venkatesh[1], John Weyant[10], Ryan Wiser[11], Alicia Zhao[6]

[1] EPRI; Palo Alto, CA, USA.
[2] Resources for the Future; Washington DC, USA.
[3] Colorado School of Mines; Golden, CO, USA.
[4] Center for Global Sustainability, University of Maryland; College Park, MD, USA.
[5] National Renewable Energy Laboratory; Golden, CO, USA.
[6] Princeton University; Princeton, NJ, USA.
[7] Rhodium Group; Oakland, CA, USA.
[8] Natural Resources Defense Council; Washington DC, USA.
[9] Center for Applied Environmental Law and Policy; Washington DC, USA.
[10] Stanford University; Stanford, CA, USA.
[11] Lawrence Berkeley National Laboratory; Berkeley, CA, USA.

*Corresponding author: John Bistline, jbistline@epri.com

**Acknowledgments**
The views and opinions expressed in this paper are those of the authors alone and do not necessarily represent those of their respective institutions, the U.S. Department of Energy (DOE), the U.S. Government, or other funding agencies, and no official endorsement should be inferred. The authors using E4ST thank Gurobi for the use of its solver and Energy Visuals Inc. for the use of its transmission data. A.F. and G.I. are also affiliated with Pacific Northwest National Laboratory, which did not provide specific support for this paper. The AAAS recognizes the U.S. Government's non-exclusive rights to use the Work for non-commercial, governmental purposes where such rights are established in the contract.

**Funding**
J.J. and Q.L. were funded by the Princeton Zero-Carbon Technology Consortium, supported by unrestricted gifts from GE, Google, and Breakthrough Energy. A.B., D.B., M.D., K.R., M.R., N.R., E.R., and D.S. were funded by RFF's Electric Power Program. M.B., A.H., and D.S. were funded by the DOE Office of Policy through the National Renewable Energy Laboratory, operated by Alliance for Sustainable Energy, LLC, for the DOE under Contract no. DE-AC36-08GO28308. R.W. was funded by the DOE Office of Policy under Lawrence Berkeley National Laboratory Contract no. DE-AC02-05CH11231.

**Competing Interests**
J.J. is part owner of DeSolve, LLC, which provides techno-economic analysis and decision support for clean energy technology ventures and investors. A list of clients can be found at https://www.linkedin.com/in/jessedjenkins. He serves on the advisory boards of Eavor Technologies Inc., a closed-loop geothermal technology company, Rondo Energy, a provider of high-temperature thermal energy storage and industrial decarbonization solutions, and Dig

Energy, a developer of low-cost drilling solutions for ground-source heat pumps and has an equity interest in each company. He also serves as a technical advisor to MUUS Climate Partners and Energy Impact Partners, both investors in early-stage climate technology companies. The other authors declare no competing interests.

**This PDF file includes:**

    Materials and Methods S1 to S7
    Figs. S1 to S19
    Tables S1 to S6
    References

**Materials and Methods**

S1: Participating Models

This model intercomparison includes nine independent energy models—six focusing on the power sector and three energy systems models. Models vary in their policy coverage (Table S1), representation of technologies (Table S2), and structure (Table S3).

Participating models include:

- **E4ST-RFF:** The [Engineering, Economic, and Environmental Simulation Tool](#) (E4ST) uses a 6,000-node model of the U.S. and Canadian grid, a linear approximation of physics-based flows, 16 frequency-weighted representative days, site-specific hourly wind and solar data, and an integrated air pollution model for comprehensive benefit-cost analysis. It represents the dispatch of system operators and the profit-maximizing generation investment and retirement decisions of investor-owners.

- **GCAM-CGS:** GCAM-USA-CGS is a version of the [Global Change Analysis Model](#) (GCAM) version 6 that is utilized and maintained by the Center for Global Sustainability. It is a multi-sector, hierarchical market equilibrium model with resolution for 50 states and D.C. GCAM-USA-CGS tracks emissions of 16 different species of air pollutants from energy, agriculture, land use, and other systems, with detailed representation of depletable primary resources and renewable sources. These energy resources are processed and consumed by end-use sectors, including buildings, transportation, and industry. The equilibrium in each period is solved by finding a set of market prices such that supplies and demands are equal in all markets. The electricity sector is divided into intermediate, subpeak, and peak load segments.

- **GenX-ZERO:** [GenX](#) is an open-source electricity resource capacity expansion model that is jointly maintained by a team of contributors at the Princeton University ZERO Lab, MIT, and NYU. GenX is a linear or mixed integer linear optimization model that determines electricity system planning at lowest cost to meet electricity demand, while subject to a variety of power system operational constraints, resource availability limits, and other imposed environmental, market design, and policy constraints.

- **Haiku-RFF:** The [Haiku](#) model is a U.S. electric sector capacity planning and system operation model. The model determines least-cost pathways to meet electricity demand, including representations of transmission, state-level markets, technology mandates, and emissions policies. The model takes as given assumptions about fuel costs, load shapes, technology costs, and demand over time.

- **IPM-EPA:** IPM-EPA results come from EPA's Regulatory Impact Analysis (RIA) for the finalized rules. EPA used the [Integrated Planning Model (IPM) 2023 reference case](#) for the RIA. IPM is an optimization model that represents the U.S. electric sector, characterized as a multi-regional, dynamic, and deterministic linear programming model. IPM meets a range of demand, transmission, environmental, dispatch, policy and reliability constraints while minimizing total system costs to determine optimal electric sector capacity expansion, dispatch, and emissions control strategies.

- **IPM-NGO:** The Integrated Planning Model (IPM) is a multi-regional, dynamic, and deterministic linear programming model of the U.S. electric power sector. IPM optimizes for the least-cost pathway for the economic construction, retirement, and operation of power plants, as well as deployment of transmission and pipelines, subject to resource adequacy requirements and environmental constraints. This version of IPM was developed for use by the Natural Resources Defense Council (NRDC) and Center for Applied Environmental Law and Policy (CAELP) and includes structural modifications and assumptions developed for and by NRDC and CAELP. IPM is a proprietary model of ICF International.

- **NEMS-RHG:** RHG-NEMS is a version of the Energy Information Administration's National Energy Modeling System (NEMS) modified by Rhodium Group. RHG-NEMS represents the US energy sector through 13 energy supply and end-use demand modules, including macroeconomic feedback and interactions with global energy markets. Spatial and temporal resolution vary across modules. The electricity module projects capacity expansion and retirement, transmission buildout, and generation required to meet future demand using a least-cost optimization while accounting for federal and state policy constraints and resource adequacy requirements.

- **ReEDS-NREL:** The Regional Energy Deployment System (ReEDS) model is an open-source capacity expansion model of the U.S. electricity sector, developed and maintained by the National Renewable Energy Laboratory (NREL). ReEDS co-optimizes generation, transmission, and energy storage assets to meet demand at the least possible cost under a range of policy and physical constraints. This version of ReEDS is largely consistent with NREL's 2024 Standard Scenarios but includes modifications in the representation of EPA's recently finalized power plant rules.

- **REGEN-EPRI:** EPRI's U.S. Regional Economy, Greenhouse Gas, and Energy (REGEN) model connects a detailed power sector capacity planning and dispatch model with energy supply and end-use modeling. Given assumptions about policies, technologies, and markets, the power sector representation determines cost-minimizing pathways for generation and energy storage investments, transmission expansion, and dispatch with a novel method of retaining temporal resolution.

Model outputs should not be interpreted as predictions, given structural and parametric uncertainties. Results in the text often compare ranges across models along with medians. Although these metrics provide indications of variation across models and central tendencies, cross-model comparisons should not be interpreted as statistical samples or probabilities of specific outcomes. All figures illustrate individual model results, and select figures show ranges to illustrate trends. In addition to showing full ranges across models, select figures also show interquartile ranges to visualize the spread of the data across models, similar to box plots, which help to identify outliers. Also, models also vary in their representations of technology-specific constraints (Table S4) and non-economic factors (Table S5), which suggests that the full uncertainty could be larger than the ranges indicate.

S2: Policy Background

The Clean Air Act (CAA) provides a federal framework for limiting emissions, establishing national standards and programs to regulate public-health-harming pollution from both stationary and mobile sources. In 2007, in *Massachusetts v. EPA*, the Supreme Court held that greenhouse gases (GHGs) should be covered under the CAA as air pollutants.[1] The EPA finalized its endangerment finding for six greenhouse gases, including $CO_2$, in 2009, stating that these GHGs threaten public health and welfare of current and future generations.[2] That finding triggered CAA obligations to issue safeguards for climate pollution, with the agency setting GHG emission limits on cars and trucks, oil and gas infrastructure, and then finalizing carbon pollution standards for existing coal plants and new gas plants in May 2024. The Inflation Reduction Act (IRA), signed into law August 2022, also reinforces the EPA's authority to regulate GHG emissions under the CAA. The law amends the CAA, expressly designating "greenhouse gases"—defined as "the air pollutants carbon dioxide, hydrofluorocarbons, methane, nitrous oxide, perfluorocarbons, and sulfur hexafluoride"—as "air pollutants."[3] These are the same substances covered by EPA's 2009 endangerment finding.[4]

---
[1] *Massachusetts v. EPA*, 549 U.S. 497, 524 (2007).

Carbon pollution standards for stationary sources like power plants are developed under Section 111 of the CAA (111 rules). The main purpose of Section 111 is to mitigate the harms to public health and welfare inflicted by air pollution from stationary sources—both existing and future—that contribute significantly to one or more air pollution problems.[5] Under Section 111(b), EPA sets performance standards for new and modified plants that immediately apply as soon a regulated plant is constructed or modified. For existing sources regulated under Section 111(d), EPA sets a standard of emissions, but unlike the standards for new sources, these standards are implemented through a state planning process. This process allows for additional flexibility based on specific determinations for individual existing plants that are unable to meet the standard (e.g., due to a facility's age and retirement date or other factors).[6] For both new and existing sources, these emission limits are based on what is "achievable" by the "best system of emission reduction" (BSER) that EPA determines is "adequately demonstrated," taking into account technical feasibility, cost, and other factors.[7] In *West Virgina v. EPA*, the Supreme Court reaffirmed that EPA has the authority to set carbon standards under Section 111, but found that the BSER should be based on technologies that "caus[e] plants to operate more cleanly" rather than the broader, generation-shifting approach that EPA had adopted in the now-vacated Clean Power Plan.[8] Critically, while the emission standards are based on the reductions that can be achieved by applying the BSER, companies are not required to use the BSER technology to comply with the emission limits. Rather, operators can meet the performance standard using other technologies, and states and utilities have the flexibility to devise their own plans to meet these standards.

In May 2024, the EPA finalized two standards—a standard on existing coal-fired electric generating units (EGUs) under Section 111(d) and a standard on new gas-fired EGUs under

---

[2] 74 Fed. Reg. 66496 (Dec. 15, 2009).
[3] IRA §60101, adding Clean Air Act §132(d)(4); §60102, adding Clean Air Act §133(d)(2); §60103, adding Clean Air Act §134(c)(2); §60107, adding Clean Air Act §135(c); §60113, adding Clean Air Act §136(i); §60114, adding Clean Air Act §137(c)(4). The same definition is included in IRA §§60105, 60106, 60108, 60111, 60112, and 60116, which appropriate funds to EPA for monitoring, reporting, and reducing GHG emissions under provisions of the Clean Air Act and other laws.
[4] U.S. EPA, Endangerment and Cause or Contribute Findings for Greenhouse Gases Under Section 202(a) of the Clean Air Act, 74 Fed. Reg. 66496 (Dec. 15, 2009).
[5] *See* 42 U.S.C. § 7411(b)(1)(A) (authorizing EPA to regulate a source if it "causes, or contributes significantly to, air pollution which may reasonably be anticipated to endanger public health or welfare").
[6] *See* 42 U.S.C. § 7411(d)(1) (permitting states to "to take into consideration, among other factors, the remaining useful life of the existing source to which such standard applies" when developing their plans).
[7] 42 U.S.C. § 7411(a)(1).
[8] *West Virginia v. EPA*, 597 U.S. 697, 706 (2022).

Section 111(b).[9] Both standards use subcategorization, with different emission rate standards based on the type of fuel, process used, and operational characteristics. Specifically, the existing coal standard is subcategorized based on expected operating life. As shown in Fig. S1, for plants expected to run beyond 2039, they must meet a standard based on 90% carbon capture by 2032. For those plants retiring before 2032, they have reporting requirements but no emission standards. And for those plants expected to run after 2032 but retire before 2039, they must meet an emission rate standard based on 40% cofiring with natural gas.

The new gas standard is subcategorized by annual utilization, or capacity factors. Peaking units—characterized by new plants operating less than 20% of the time—have a lower-emitting fuels requirement equivalent to a limit of 160 pounds of $CO_2$ per million British thermal units (MMBtu). Intermediate load units—those operating between 20% and around 40% of the time—will need to meet a limit based on a highly-efficient simple-cycle turbine (1,150 lbs/MWh limit). And baseload new gas plants—those operating above 40%—will need to meet a limit based on 90% carbon capture by 2032.

EPA's proposed 111 rules from May 2023 also included standards for existing gas-fired units; however, these standards were not finalized with the other two carbon standards. In February 2024, EPA announced that it would take additional time to finalize these standards to develop a more comprehensive approach that would cover the entire fleet of existing gas-fired turbines, which also would address other toxic and criteria air pollutants.[10]

S3: Scenario Design

To evaluate impacts on emissions and power sector outcomes, scenarios with the finalized 111 rules are compared to their counterfactual reference scenarios without these rules. The reference scenario includes other federal and state policies and incentives as of early 2024. On-the-books policies are summarized in Table S1, including IRA provisions, Infrastructure Investment and Jobs Act (IIJA), other finalized EPA rules,[11] as well as state-level policies and standards. Models include major power sector IRA provisions such as the 45Y production tax credit, 48E investment tax credit, 45Q $CO_2$ capture and storage credit, and 45V clean hydrogen production credit. Power sector production and investment tax credits generally include scenario-specific expirations starting when power sector $CO_2$ emissions reach 25% of their 2022 levels.

Section 111 scenarios assume unit-level rate-based compliance without trading across power plants or states. These rate-based standards include flexibility about the subcategory choice for existing coal and new gas units and compliance. These scenarios do not include potential remaining useful life and other factors (RULOF) exemptions or other flexibilities, such as trading across units or regions.

To the extent that model structures allow, the following assumptions were harmonized across models and are held constant across scenarios:

---

[9] 89 Fed. Reg. 39798 (May 9, 2024).
[10] "Statement from EPA Administrator Michael S. Regan on EPA's approach to the power sector," *EPA Press Release*, February 29, 2024, https://www.epa.gov/newsreleases/statement-epa-administrator-michael-s-regan-epas-approach-power-sector.
[11] Along with the final rules for $CO_2$ emissions from existing coal-fired and new gas-fired power plants, EPA finalized a suite of rules for power plants, including updated Mercury and Air Toxics Standards (MATS), wastewater discharge standards, and regulations for coal combustion residuals. Most models represent these rules in all scenarios (Table S1).

- **Technological costs:** Capital costs come from NREL's 2023 Electricity Annual Technology Baseline (ATB) Technologies "Moderate" scenario (Fig. S2).
- **Financing assumptions:** Models assume weighted average cost of capital (WACC) or discount rate assumptions of 6% (real) for electric sector investments.
- **Fuel costs:** Models assume the U.S. EIA's 2023 Annual Energy Outlook "Reference" scenario (Fig. S3).
- **Planning reserve margin:** Models assume planning reserve margins of 15% for each region.

Reference and 111 scenarios are run with reference load growth and separately with higher electricity demand. Reference load scenarios exhibit growth of 18-34% by 2035 from 2022 levels across models (Fig. S16), and the higher load sensitivity increases annual values by an additional 3.3% over the reference in 2030, 7.3% in 2035, and 10.7% in 2040, which are based on EPA's "High Demand" sensitivity in the rules RIA (Fig. S16).[12] These increases are applied to all regions and intra-annual segments, which preserves regional variation and load shapes in models.[13]

To understand potential impacts of future existing source performance standards for existing natural gas power plants, we also conduct a sensitivity that adds coverage of existing gas to the rules. This stylized scenario adopts similar subcategories and timing as the finalized new source standard for gas (Fig. S1):

- Existing gas-fired plants with anticipated capacity factors above 40% would have standards based on the application of CCS with 90% capture rates by 2035.
- Units with capacity factors less than 40% have less stringent requirements and can operate without additional controls if they use "highly efficient generation" or "low-emitting fuels."

This scenario does not include potential RULOF exemptions or subcategory-specific exemptions based on unit characteristics (e.g., size, which was a feature of the proposed existing source standards in May 2023), which makes this scenario an upper bound on impacts. Note that details about potential emissions standards for existing gas are highly uncertain, given contrasting feedback from stakeholders as part of EPA's non-regulatory docket in early 2024.

Models vary in the varieties of emissions control options represented for existing and new fossil fuel capacity (Table S2). Models use native load growth assumptions in the reference case but broadly align in electricity demand (Fig. S16).

---

[12] Note that the stated motivation in the RIA is to account for electricity demand associated with EPA's finalized vehicle rules as well as the non-transport load from the Annual Energy Outlook 2023 "High Economic Growth" scenario. However, since many participating models in this study account for these rules in their reference demand scenarios, these sensitivities account for higher-than-expected demand from data centers, manufacturing, and end-use electrification but do not attempt to represent potential regional distributions of these new loads.

[13] Hourly load shapes and the regional allocation of electricity demand growth likely vary based on the composition of these loads. For instance, data centers could have relatively flat load shapes, and their growth has historically been concentrated in several U.S. states [11].

S4: Additional Emissions Results

Model results suggest that the rules reduce $CO_2$ by 120-390 Mt-$CO_2$ annually by 2035 compared to current policies without rules (Fig. S4). Incremental impacts of the rules exhibit U-shaped effects for many models, with the largest impacts in 2035 and smaller effects in 2030 (due to fewer requirements under the rules by that period) and in 2040 (since reference conditions include more coal retirements over time and more low-emitting electricity deployment before IRA credits expire). Emissions reductions induced by the rules are lower for models with lower reference $CO_2$ and coal generation (and vice versa).

Emissions by source change over time, which historically were dominated by coal but increasingly come from gas-fired generation (Fig. S5). New and existing gas-fired generation accounts for 49% of power sector $CO_2$ in 2023 but is expected to be 55-79% of 2035 emissions under reference conditions.

EPA's power plant rules also lower co-pollutants, including $NO_x$ and $SO_2$, though these emissions are already expected to continue their downward trend with current policies (Fig. S6). Note that this figure shows reductions from 2015 levels, instead of 2005 like the other figures, since declines from 2005 have been significant enough to make it difficult to visualize differences in future emissions.

S5: Additional Power Sector Results

Before accounting for the rules, projected coal retirements are consistent with historical market trends (Fig. S7), which have seen steady coal plant closures since 2011, and are typically higher than announced plant retirements. However, this analysis suggests that the phase-down of unabated coal is accelerated and deepened with EPA's power plant rules.

Historical capacity additions by technology can be compared with projected investments with and without the rules (Fig. S8). The rules do not require widespread CCS deployment, even with IRA incentives and state-level policy, with 0-18 GW of coal retrofits and 1-24 GW of CCS-equipped gas by 2040 under the rules.

There is also cross-model variation in capacity and generation responses to the rules (Fig. S9) and capacity factors (Fig. S11), which differ over time and by technology.

S6: Additional Cost Results

Model results indicate relatively low incremental costs of the rules, which are 0.5-3.7% of system costs through 2050 (Fig. S14), despite cross-model variation in the composition and extent of power system expenditures. Differences across models are connected to the alternate timing and degree of capacity investments (Fig. S10) and to the scope of included costs. These low costs are also reflected in the small wholesale electricity price impacts relative to the reference (Fig. 2 and Table S6).[14]

---

[14] Note that wholesale electricity price changes are reported and not retail price impacts. Retail prices are higher than wholesale prices and include transmission and distribution costs, and since these delivery categories likely do not vary across scenarios by as much as the generation component of retail prices, retail price impacts of the 111 rules are generally smaller than wholesale electricity price effects.

These estimates of changes in total system costs can be used with total emissions changes to calculate average abatement costs across models (Fig. S15).[15] Average abatement costs range from $6-44/t-$CO_2$, which indicates that the climate benefits alone of the rules likely exceed costs by a large margin. Fig. S15 compares abatement costs with ranges of social cost of carbon values in the literature and EPA estimates, but recent work suggests that these values could be higher with alternate assumptions about equity weighting [1] and updated expert assessments of model structure and discounting [2]. This finding reflects the relatively low abatement cost of coal switching, which is consistent with the decarbonization literature [3] . Abatement costs of the rules are also lower than the cost of controls for other EPA rules[16] (dotted line in Fig. S15) and costs of $CO_2$ reduced for BSER options in the rules.

S7: Additional Scenario Results

Total electricity demand increases by 18-34% in the reference demand scenario by 2035 relative to 2022 levels and 27-44% in the higher demand case. Generation and emissions responses to these higher demand levels are shown in Fig. S17. Note that the natural gas responses in these scenarios are not only increased dispatch of existing resources but also new builds for some models.

Several additional uncertainties could alter electric sector decisions, which have been explored elsewhere in the context of EPA's power plant rules:

- Natural gas prices: Three models investigate how lower or higher natural gas prices could change electric sector $CO_2$ emissions and impacts of EPA's power plant rules [4, 5, 6]. Although these studies suggest that long-run electric $CO_2$ is higher with lower natural gas prices (and lower $CO_2$ with higher prices), the incremental impact of the rules under alternate fuel price scenarios is ambiguous. For instance, under lower natural gas prices, GenX-ZERO finds lower $CO_2$ reductions from the final rules relative to reference gas prices [4], EPA-IPM also finds lower $CO_2$ reductions [5], and NEMS-RHG finds higher $CO_2$ reductions [6].[17]

- Technology costs: Capital costs are uncertain not only for emerging technologies (e.g., enhanced geothermal, advanced nuclear, CCS) but also for technologies that have seen considerable cost declines and deployment in recent years (e.g., solar, battery storage, wind). NEMS-RHG varies solar, land-based/offshore wind, battery storage, and CCS-equipped gas costs across reference and 111 scenarios [6], though other input assumptions are varied simultaneously, which makes it challenging to isolate changes from technological cost assumptions. GenX-ZERO conducts sensitivities with conservative and optimistic wind and solar assumptions, and although $CO_2$ is lower with optimistic assumptions (and higher with conservative assumptions), the incremental $CO_2$ emissions impact of the rules is similar across technology sensitivities [4].

---

[15] Average abatement costs are calculated by dividing the change in net present value of system costs between the reference and 111 scenarios across the time horizon by cumulative abatement. Although marginal abatement costs are typically compared across policies and mitigation options [4], one feature of regulatory approaches like the performance standards in this analysis is that they mask marginal costs, and model implementations of the rules do not lend themselves to simple calculations of marginal abatement costs. This differs from $CO_2$ caps or other climate policy instruments where shadow prices on $CO_2$ constraints can be straightforward to calculate and provide an economic interpretation [5].

[16] In the finalized 111 rules, EPA compares costs to GHG controls in rulemakings for other industries. The metric for cost reasonableness is based on the 2016 new source performance standards regulating GHGs for the Crude Oil and Natural Gas source category, where EPA found $2,447/ton to be reasonable for methane reductions, which is $98/ton of $CO_2$e reduced.

[17] Note that these studies assume different natural gas price trajectories for their lower and higher scenarios and, in their individual studies, may make different assumptions from the harmonized scenarios in this analysis. REGEN-EPRI finds similar incremental $CO_2$ reductions from the proposed rules under reference and higher natural gas prices [10].

- Alternate capacity factor thresholds: IPM-NGO runs scenarios with alternate capacity factor thresholds for existing and new gas-fired plants in an analysis of EPA's proposed power plant rules and finds slightly lower $CO_2$ emissions than scenarios with higher thresholds [7].

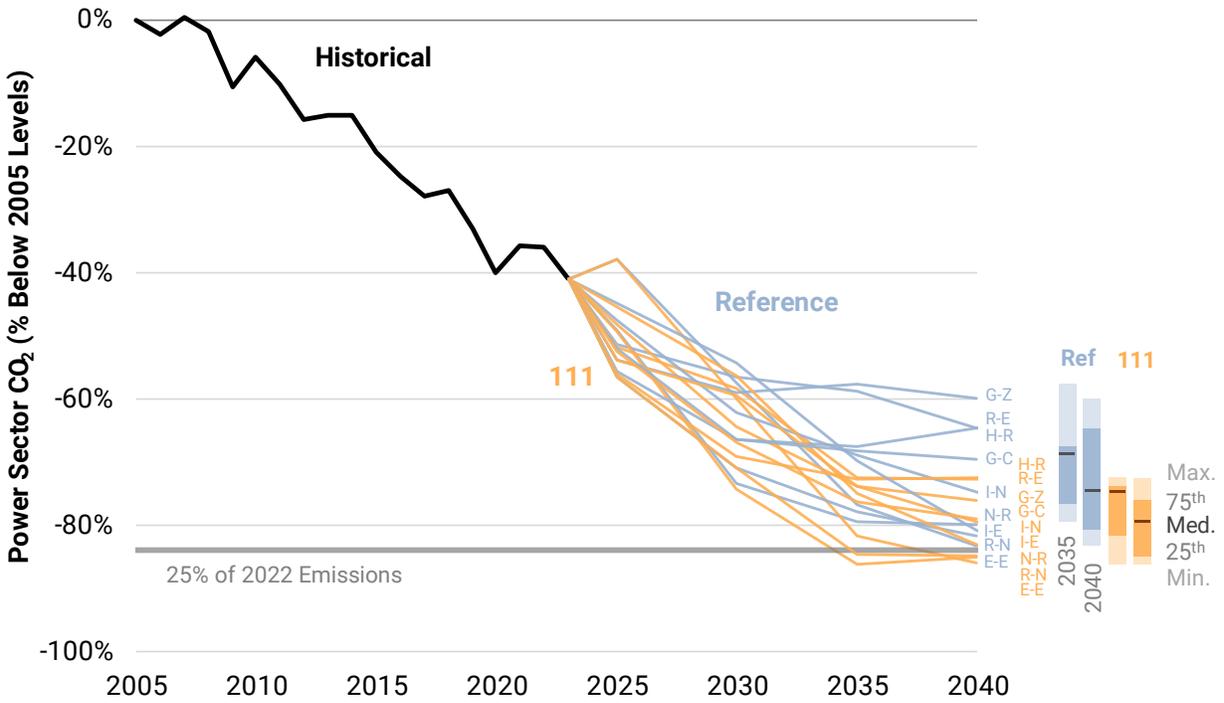

**Fig. 1. Cross-model comparison of U.S. power sector emissions reductions over time with and without EPA's power plant rules.** Ranges on the right show minimum, 25th percentile, median, 75th percentile, and maximum values across models with current policies only ("Reference") and with EPA rules ("111"). Inflation Reduction Act tax credits begin to expire in 2032 or after power sector $CO_2$ reaches 25% of 2022 levels, whichever is later. Details on models and study assumptions are provided in Materials and Methods S1 and S3.

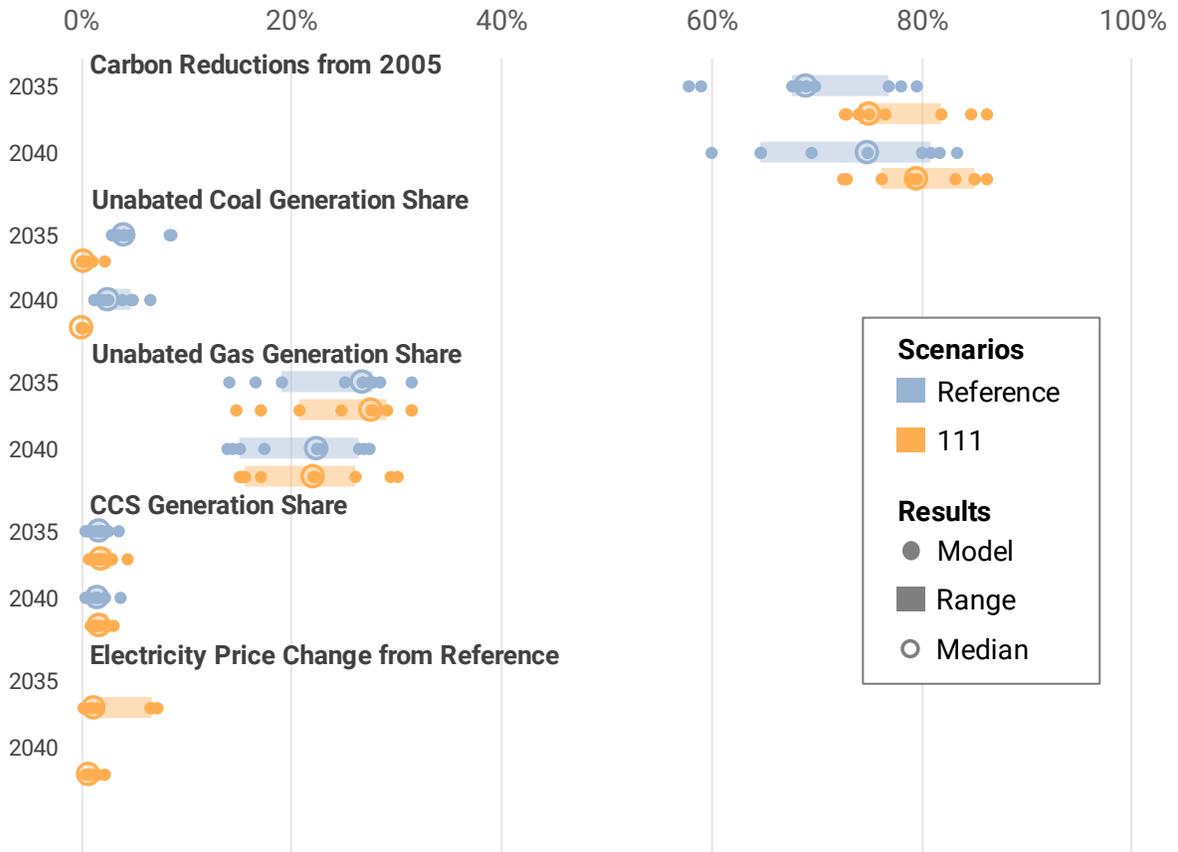

**Fig. 2. Summary of key indicators for 111 and reference scenarios across models.** From the top, indicators are electric sector $CO_2$ reductions (% from 2005 levels), generation share from coal without CCS, generation share from natural gas without CCS, CCS-equipped generation share, and wholesale electricity price changes relative to the reference scenario. For each metric and scenario, values are shown for individual models (dots), median (open circles), and interquartile range across models (bars).

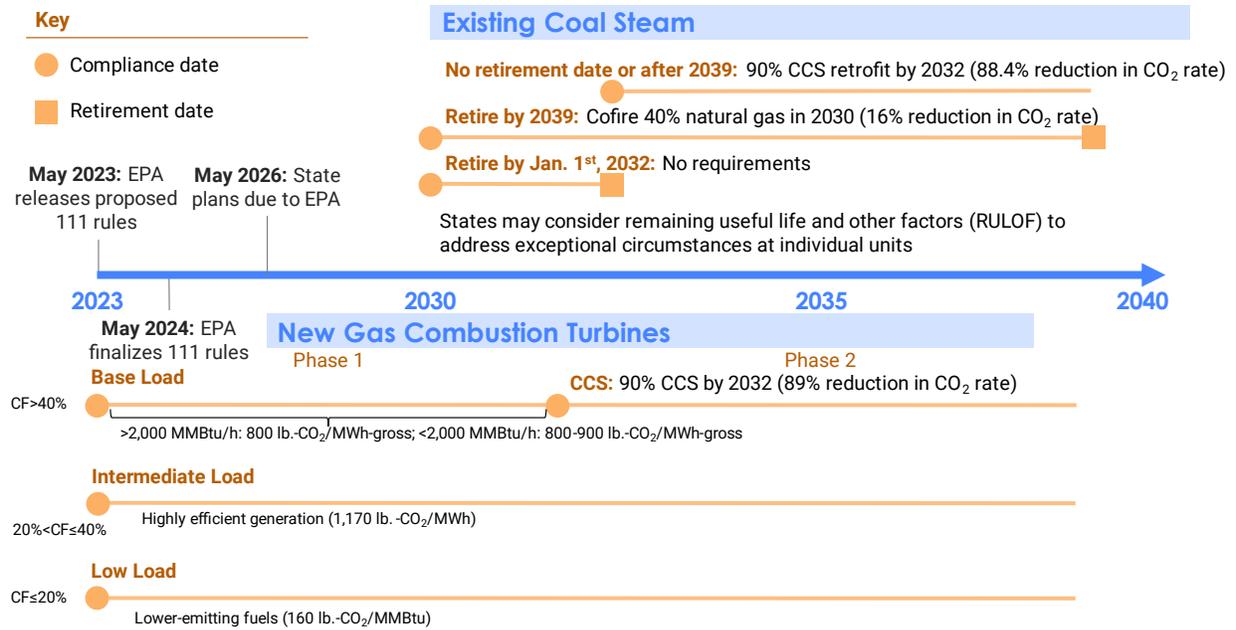

**Fig. S1. Timing and subcategories for EPA's finalized 111 power plant rules.** Existing source performance standards for coal plants under Section 111(d) of the Clean Air Act are shown at the top, and new source performance standards for gas combustion turbines under 111(b) are shown on the bottom. New gas combustion turbines are defined as any combustion turbine EGUs that commenced construction or reconstruction after May 23, 2023.

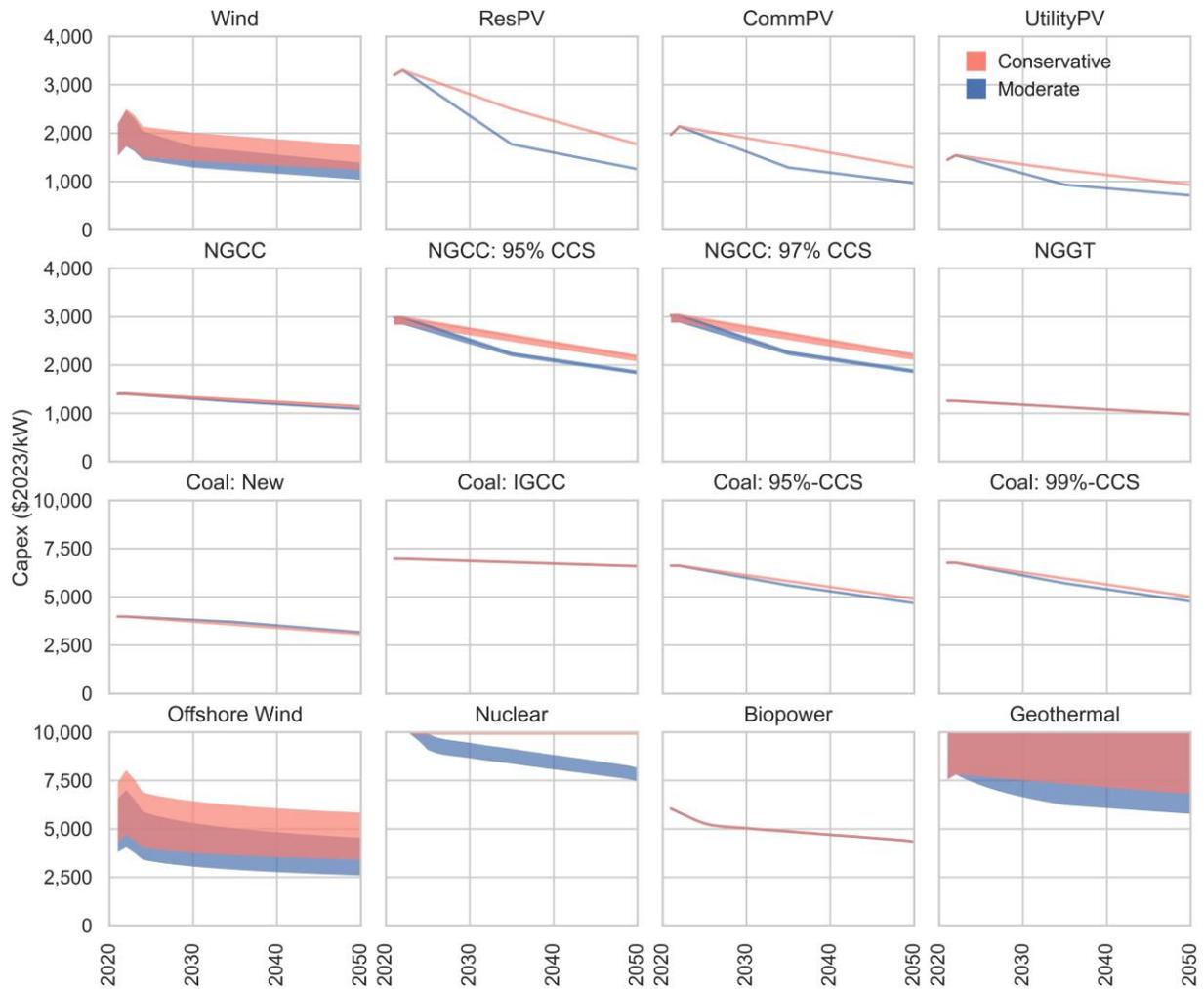

**Fig. S2. Capital cost assumptions of key power sector technologies over time by model.**
Costs are shown in 2023 U.S. dollars per kilowatt capacity (solar capacity in kW$_{AC}$). Assumptions are based on NREL's 2023 Electricity ATB Technologies ([link](#)). CCS = carbon capture and sequestration; Comm = commercial; NGCC = natural gas combined cycle; NGGT = gas turbines; PV = photovoltaic; Res = residential.

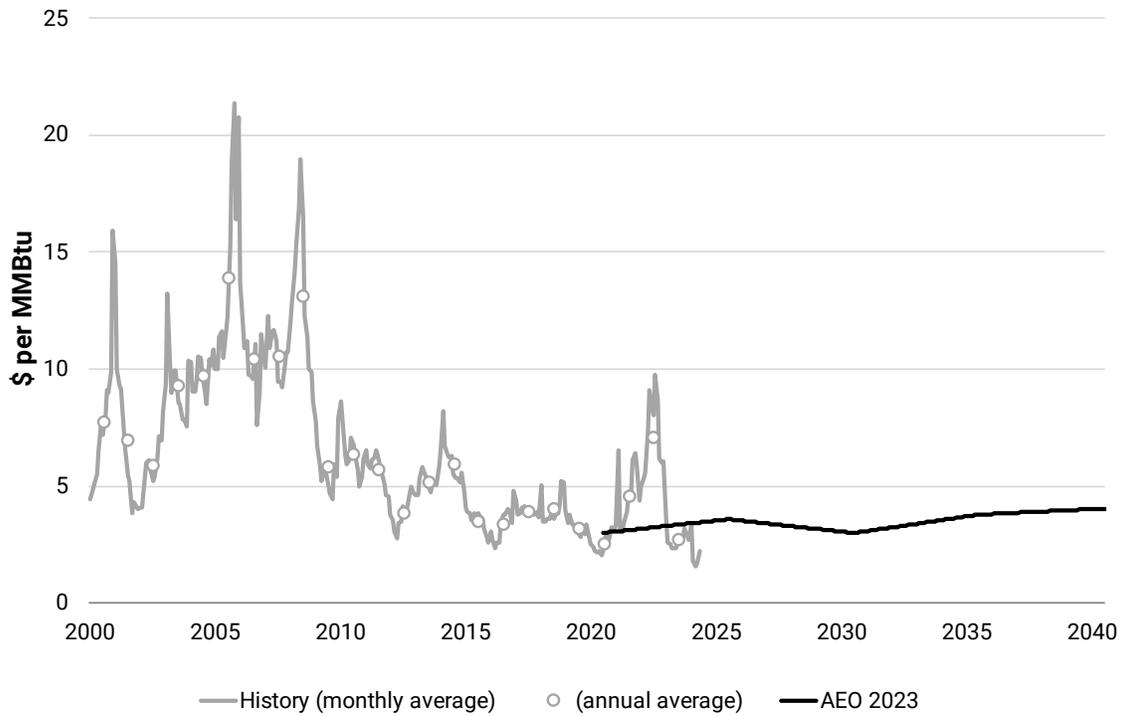

**Fig. S3. Historical and projected natural gas price assumptions over time.** Values are shown in real 2023 U.S. dollar terms. Projections come from the U.S. EIA's 2023 Annual Energy Outlook "Reference" scenario.

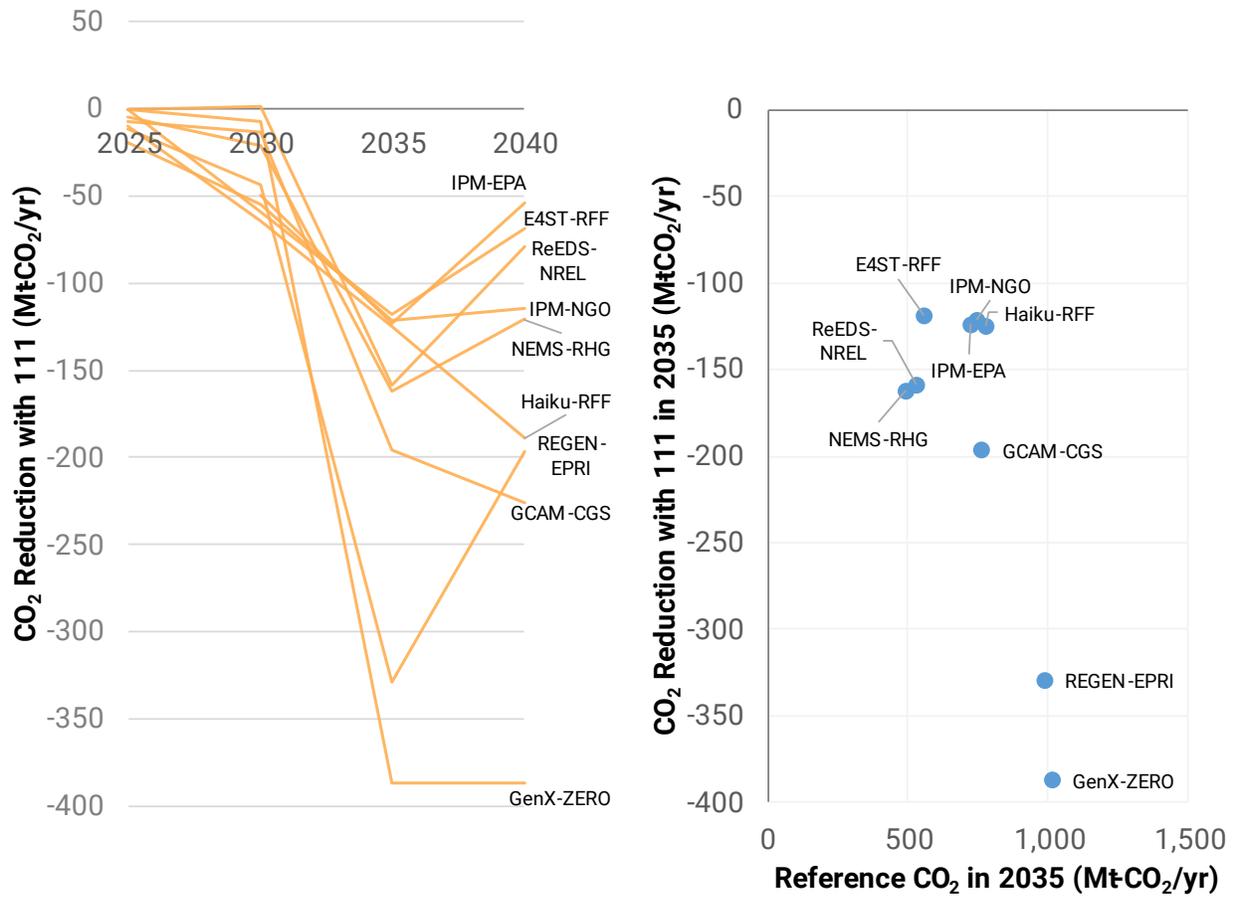

**Fig. S4. Incremental impacts of 111 rules on CO₂ by model.** (**A**) Difference between reference and 111 scenarios in absolute terms for each participating model (million metric tons of $CO_2$). (**B**) Scatter plot of incremental emissions reduction with rules versus emissions in the reference without the rules.

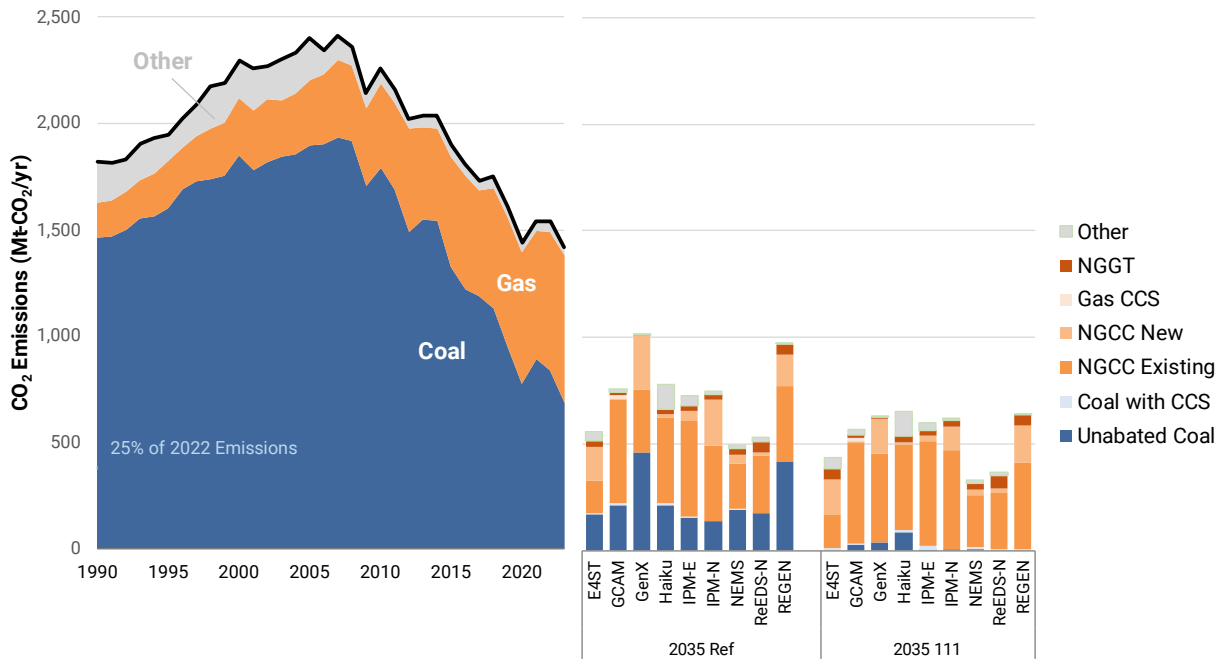

**Fig. S5. CO$_2$ emissions by asset type and model for scenarios without and with EPA's power plant rules (Reference and 111, respectively).** Historical emissions come from U.S. EPA's Inventory of U.S. Greenhouse Gas Emissions and Sinks. CCS = carbon capture and sequestration; NGCC = natural gas combined cycle; NGGT = gas turbines (including all non-NGCC gas capacity).

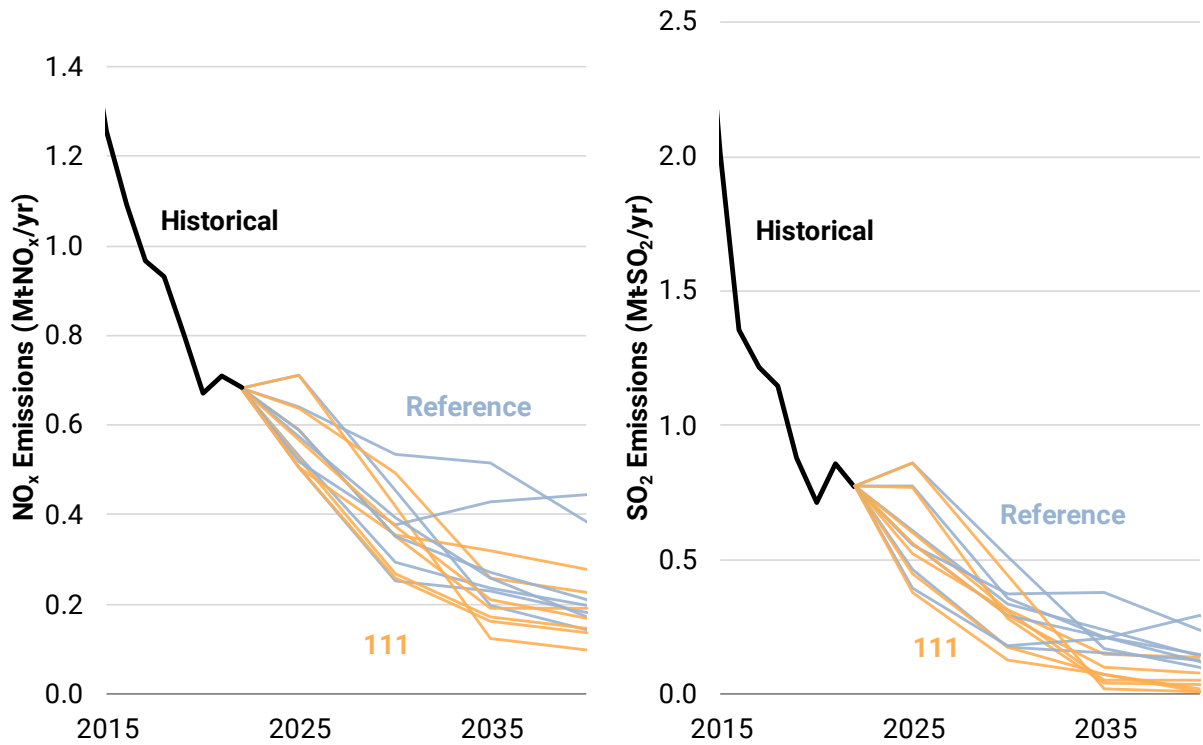

**Fig. S6. NO$_x$ and SO$_2$ emissions by model for scenarios without and with EPA's power plant rules (Reference and 111, respectively).** Historical emissions come from U.S. EPA's Progress Report on Emissions Reductions.

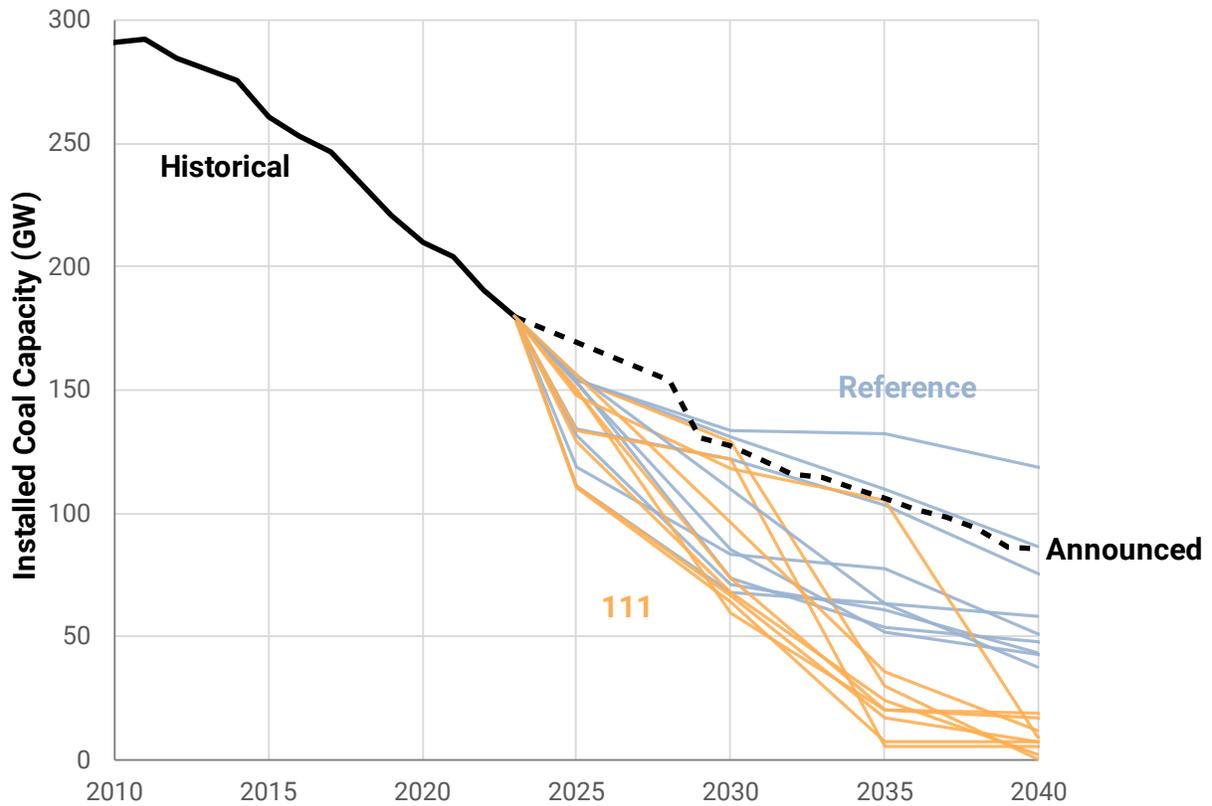

**Fig. S7. Historical and projected installed coal capacity by model, including unabated, cofiring, and CCS-equipped capacity.** Historical values come from U.S. EIA's Form EIA-860 data. Announced retirements are based on U.S. EPA's National Electric Energy Data System (NEEDS) database from January 2024.

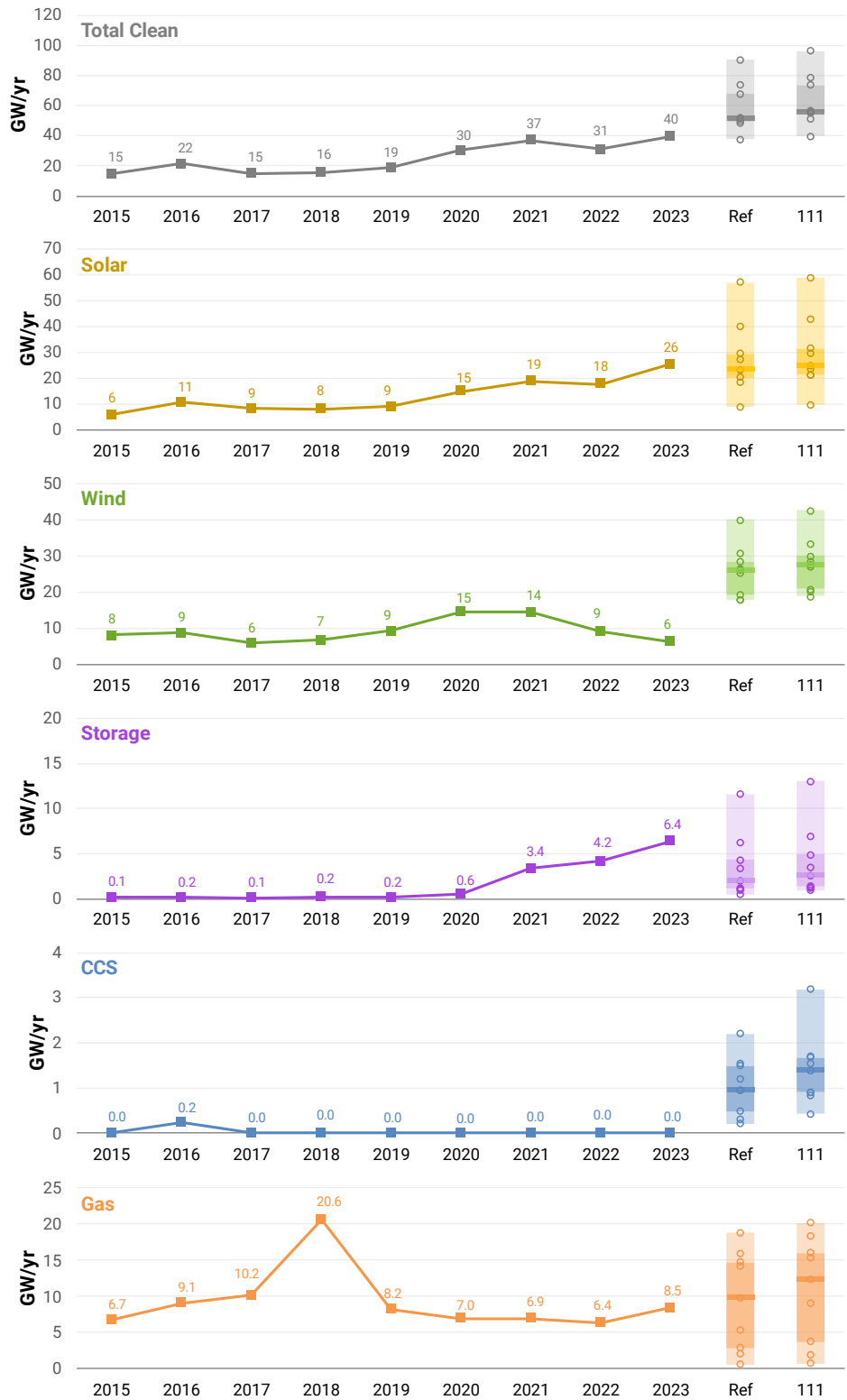

**Fig. S8. Historical and projected capacity additions by technology.** Projections show average annual additions through 2035 under a reference scenario without the power plant rules (Ref) and with EPA's rules (111). Ranges across model projections show the difference between the minimum and maximum values (light), interquartile range (middle), and median (dark). Historical values come from U.S. EIA's Form EIA-860 data through 2023. Clean capacity in top panel includes renewables, energy storage, nuclear, and CCS-equipped capacity. Solar capacity includes both distributed and non-distributed solar (in $GW_{AC}$ terms). CCS panel includes CCS-equipped coal, natural gas, and biomass.

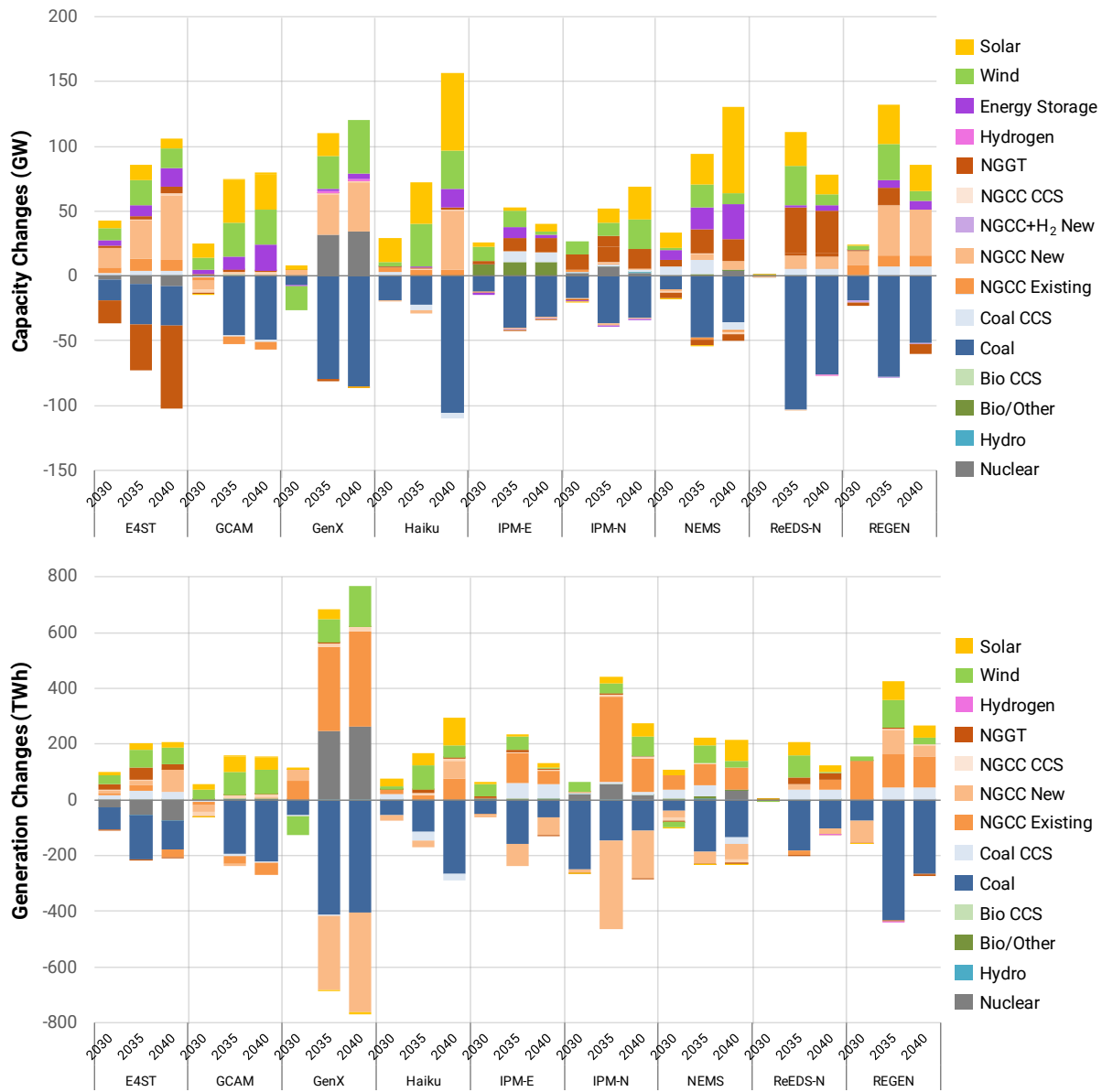

**Fig. S9. Differences in projected capacity and generation with 111 rules (relative to the reference scenario) by technology and model. (A)** Capacity changes, where solar is shown in $GW_{AC}$ terms. **(B)** Generation changes. CCS = carbon capture and sequestration; NGCC = natural gas combined cycle; NGGT = gas turbines (including all non-NGCC gas capacity).

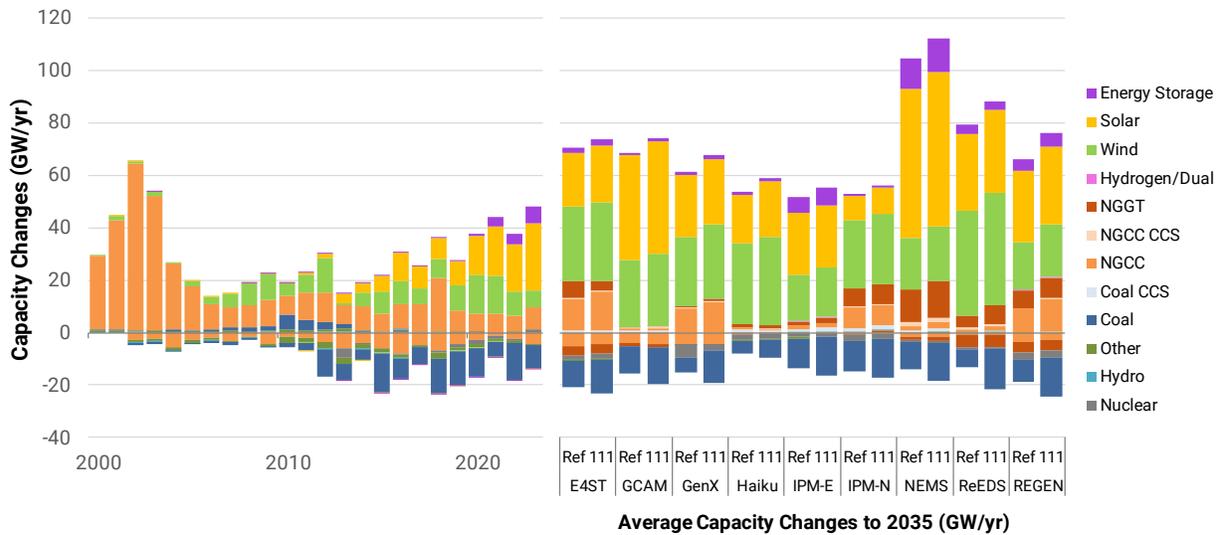

**Fig. S10. Electricity investments and retirements by generation and energy storage technology across models.** Historical values are based on EIA-860 data through 2023, and modeled results show average annual changes through 2035 without and with EPA's power plant rules ("Ref" and "111," respectively). CCS-equipped NGCC capacity is typically new builds, while CCS-equipped coal capacity is retrofits. CCS = carbon capture and sequestration; NGCC = natural gas combined cycle; NGGT = gas turbines (including all non-NGCC gas capacity).

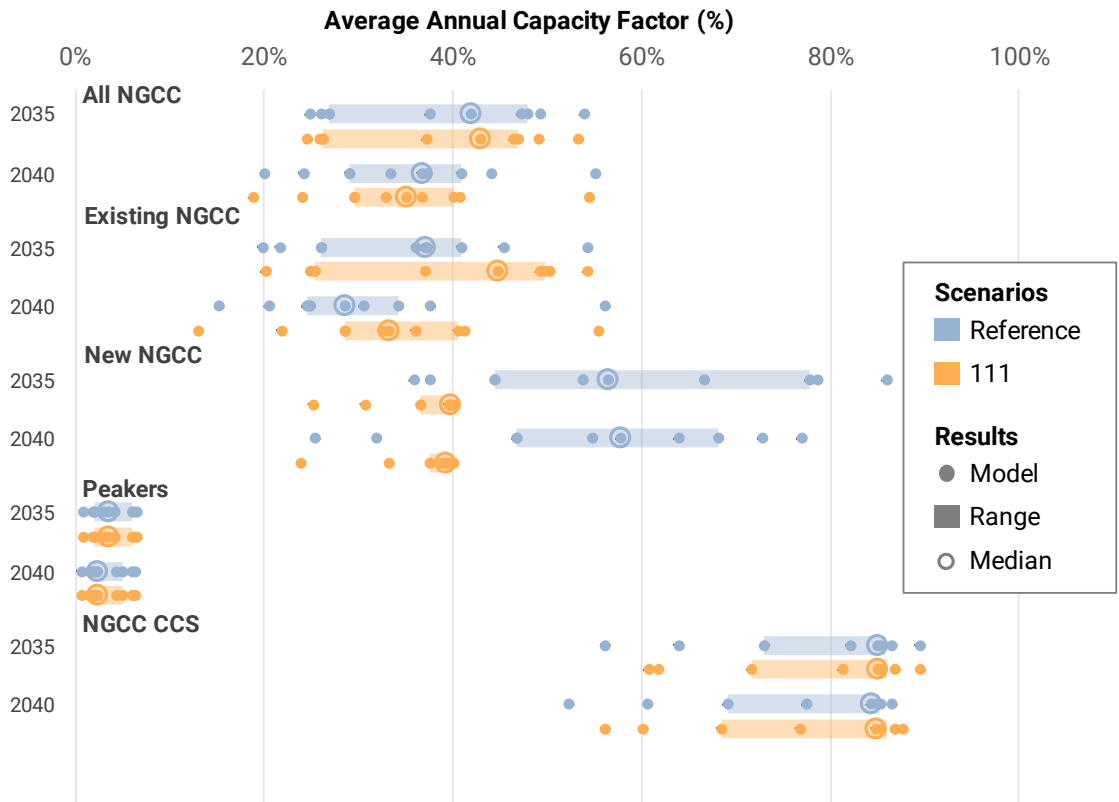

**Fig. S11. Capacity factors for gas-fired assets by model, scenario, and time period.** Dots show individual model results, bars show the interquartile range, and circles show the median across models. CCS = carbon capture and sequestration; NGCC = natural gas combined cycle.

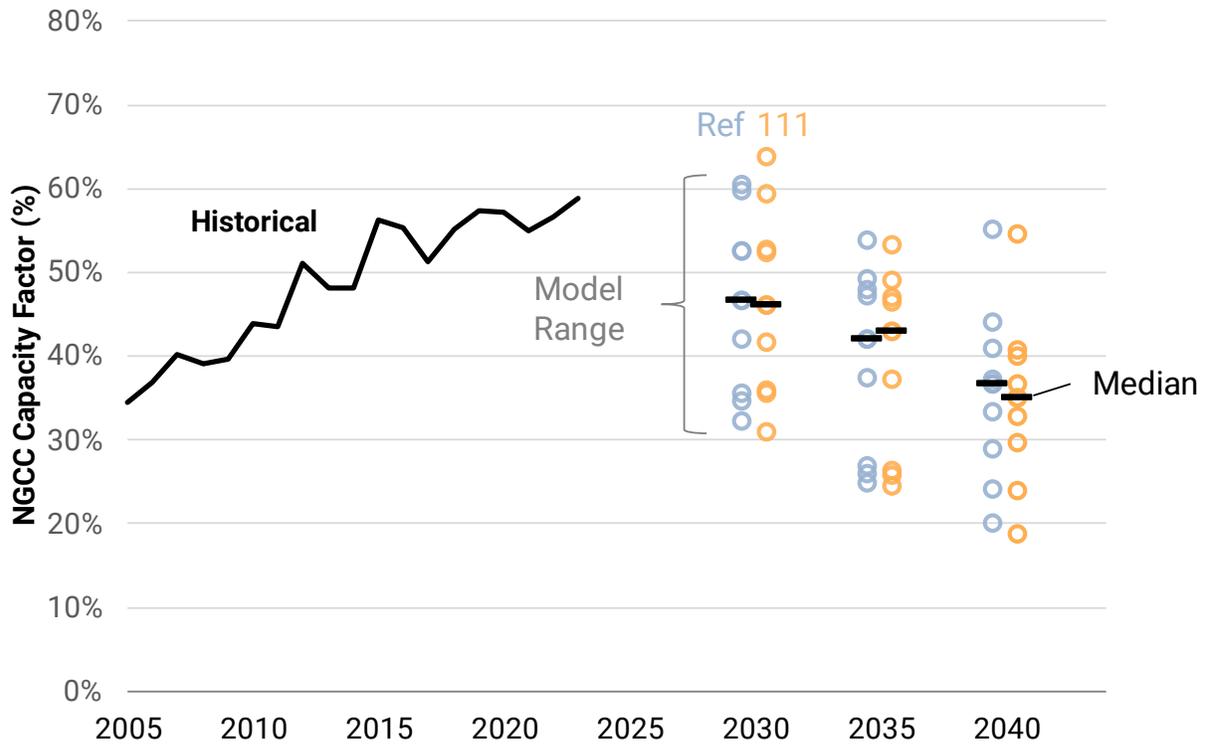

**Fig. S12. Capacity factors for new and existing natural gas combined cycle capacity by scenario and model.** Circles show individual model results, and dashes show the median capacity factors across models.

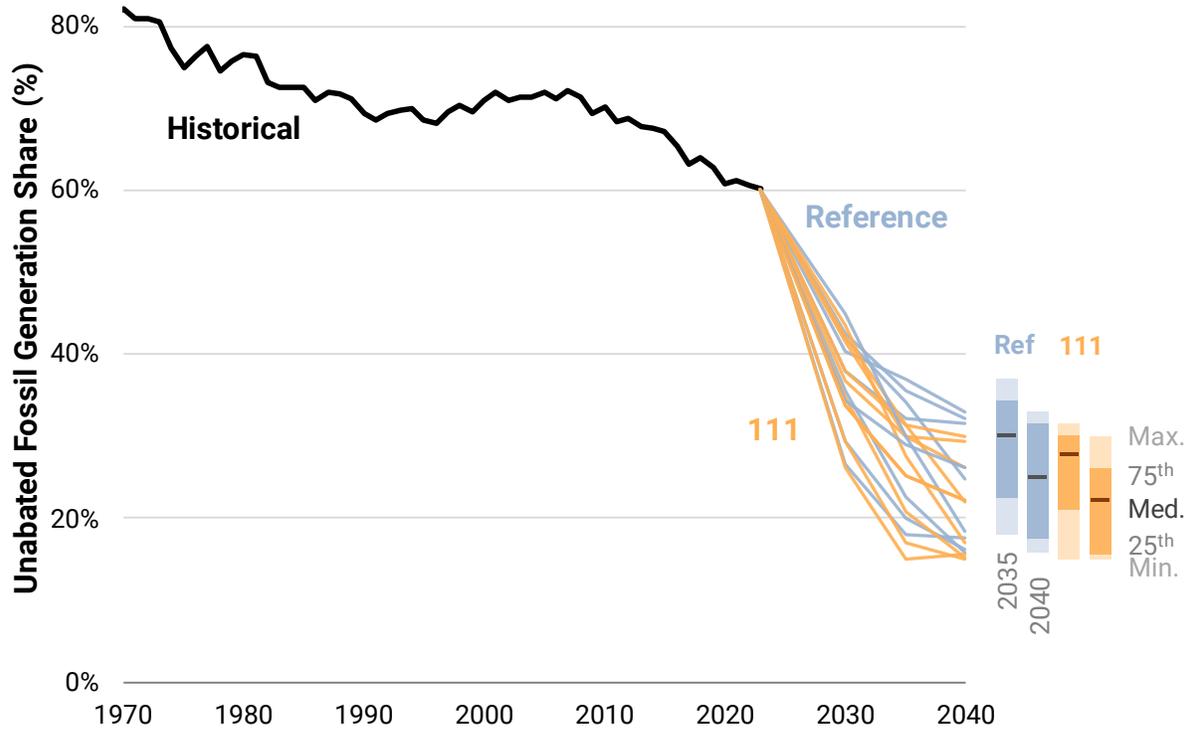

**Fig. S13. Fossil fuel generation share without carbon capture by scenario and model.** Ranges on the right show minimum, 25[th] percentile, median, 75[th] percentile, and maximum values across models. Historical data are from U.S. EIA's Monthly Energy Review.

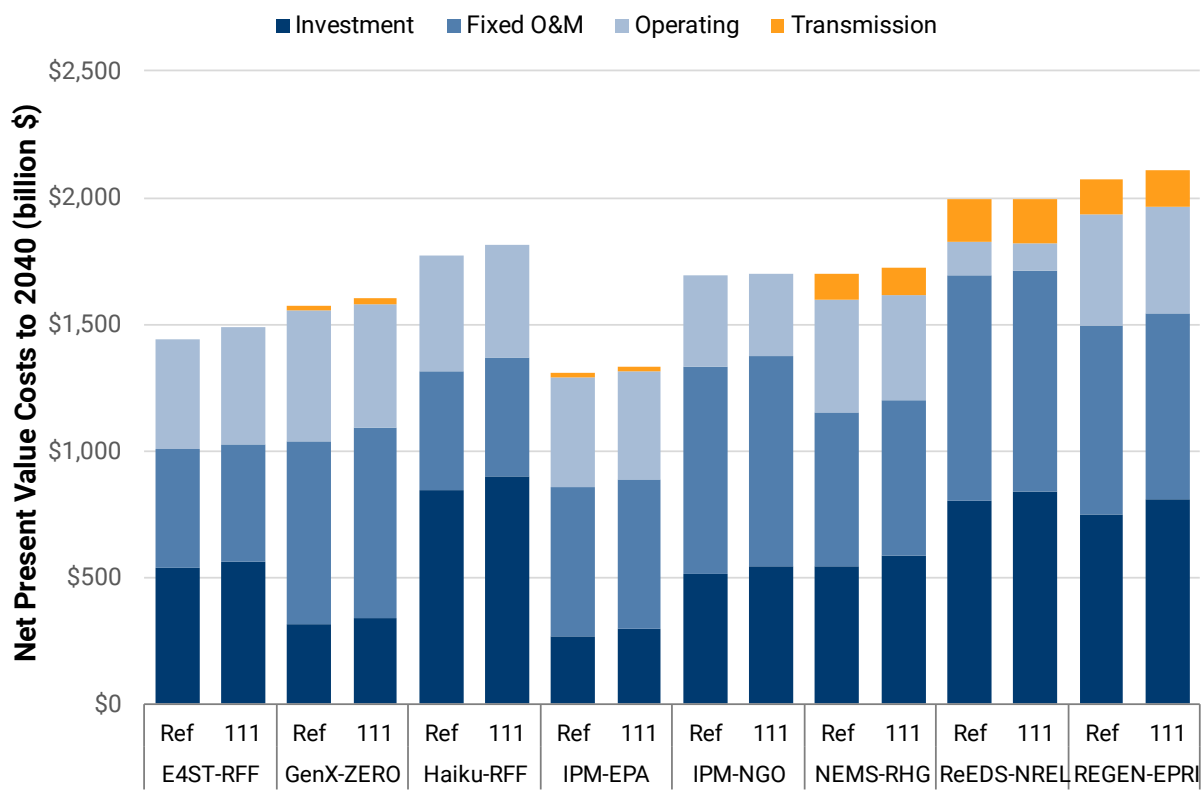

**Fig. S14. Net present value of power system costs by scenario and model through 2040.**
Costs are disaggregated into generation and energy storage investments, fixed operations and maintenance (O&M) costs (e.g., labor, maintenance contracts, insurance), operations-related expenditures (e.g., fuel costs, variable O&M costs), and interregional transmission costs (i.e., not including intraregional transmission or distribution costs). Values are shown in real 2023 U.S. dollar terms discounted at 6% across the reported time horizons of the models. Costs are reported for 2025 onwards for most models, except EPA-IPM which reports costs starting in 2030.

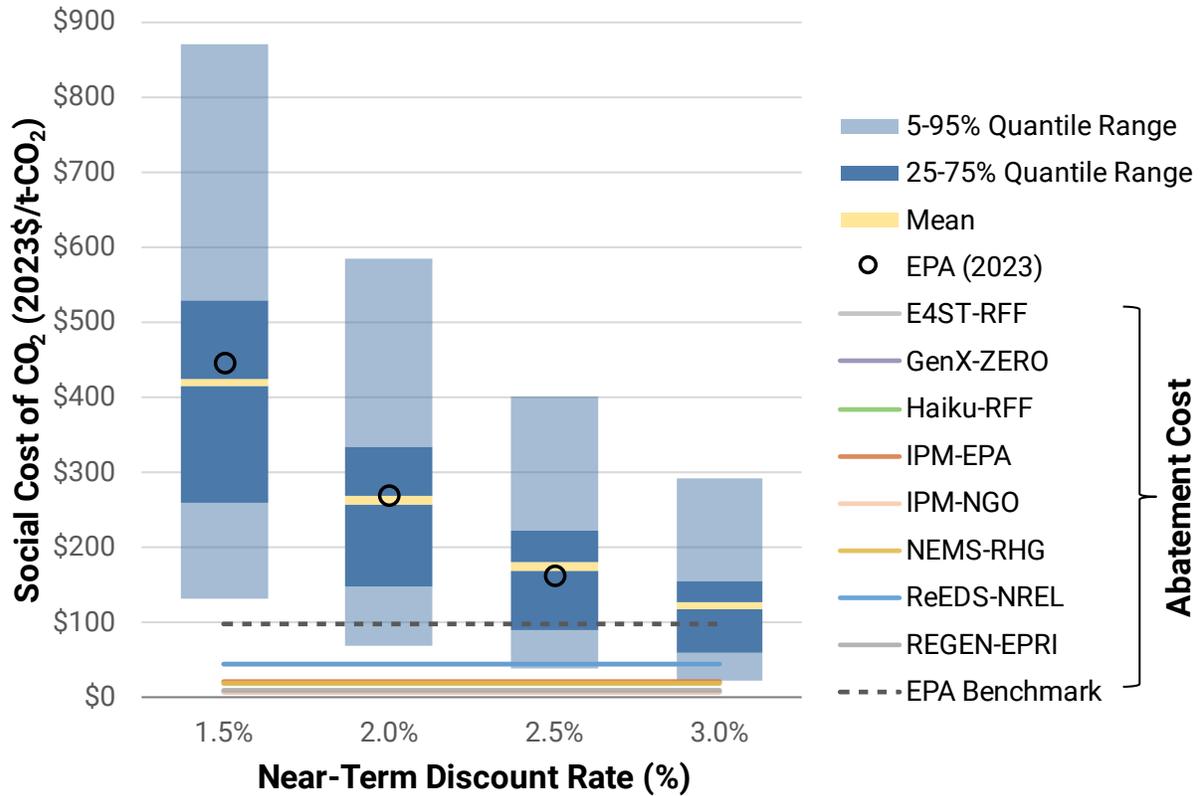

**Fig. S15. Comparison of average abatement costs of EPA power plant rules by model and recent estimates for the social cost of carbon in 2030.** Quantile range and mean social cost of carbon values come from Rennert, et al. (2022) across near-term discount rate assumptions, and circles show values from EPA (2023), which are used in the RIA for the 111 rules. The dotted line shows costs of emissions controls used for comparison in the 111 RIA, which is based on 2016 new source performance standards regulating GHG in the Crude Oil and Natural Gas source category. All values are shown in real 2023 U.S. dollar terms.

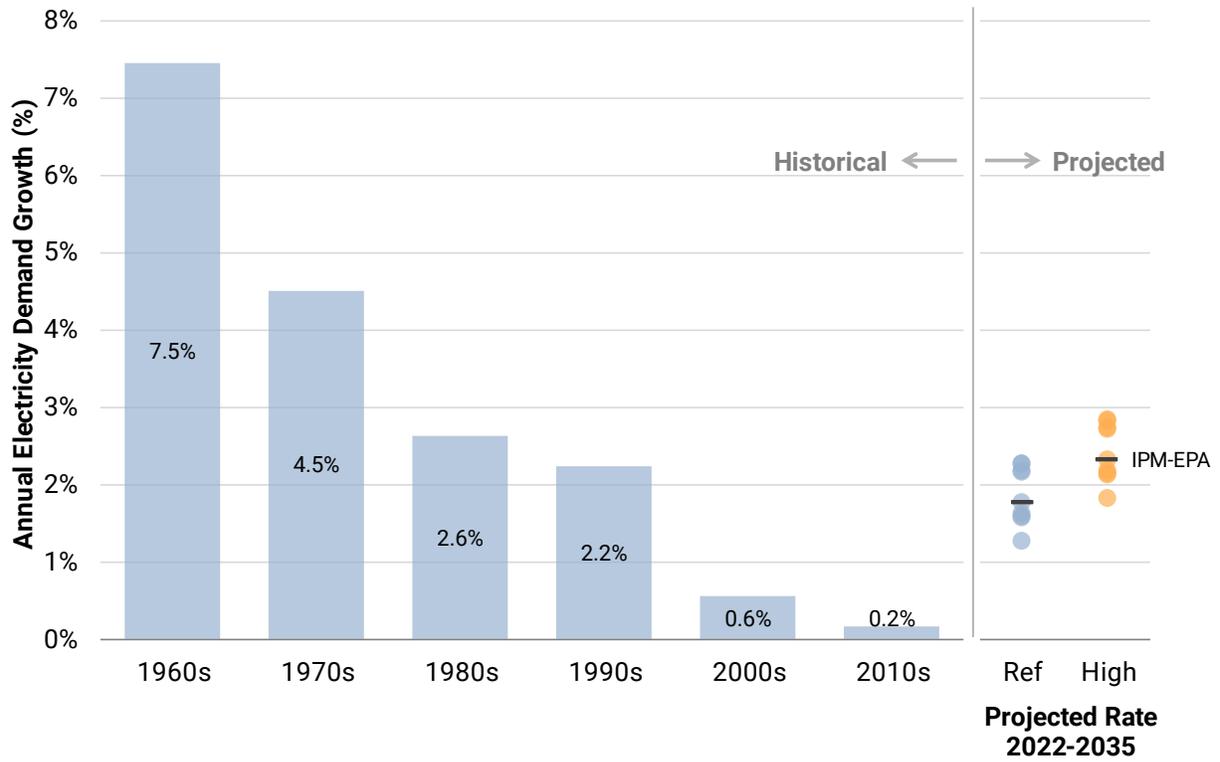

**Fig. S16. Historical and projected growth rates of annual electricity demand.** Compound annual growth rates are shown. Projections show model-specific values from 2022-2035 across the reference and high load growth scenarios. IPM-EPA results come from EPA's Regulatory Impact Analysis for the finalized 111 rules. Historical values come from U.S. EIA's State Energy Data System (SEDS).

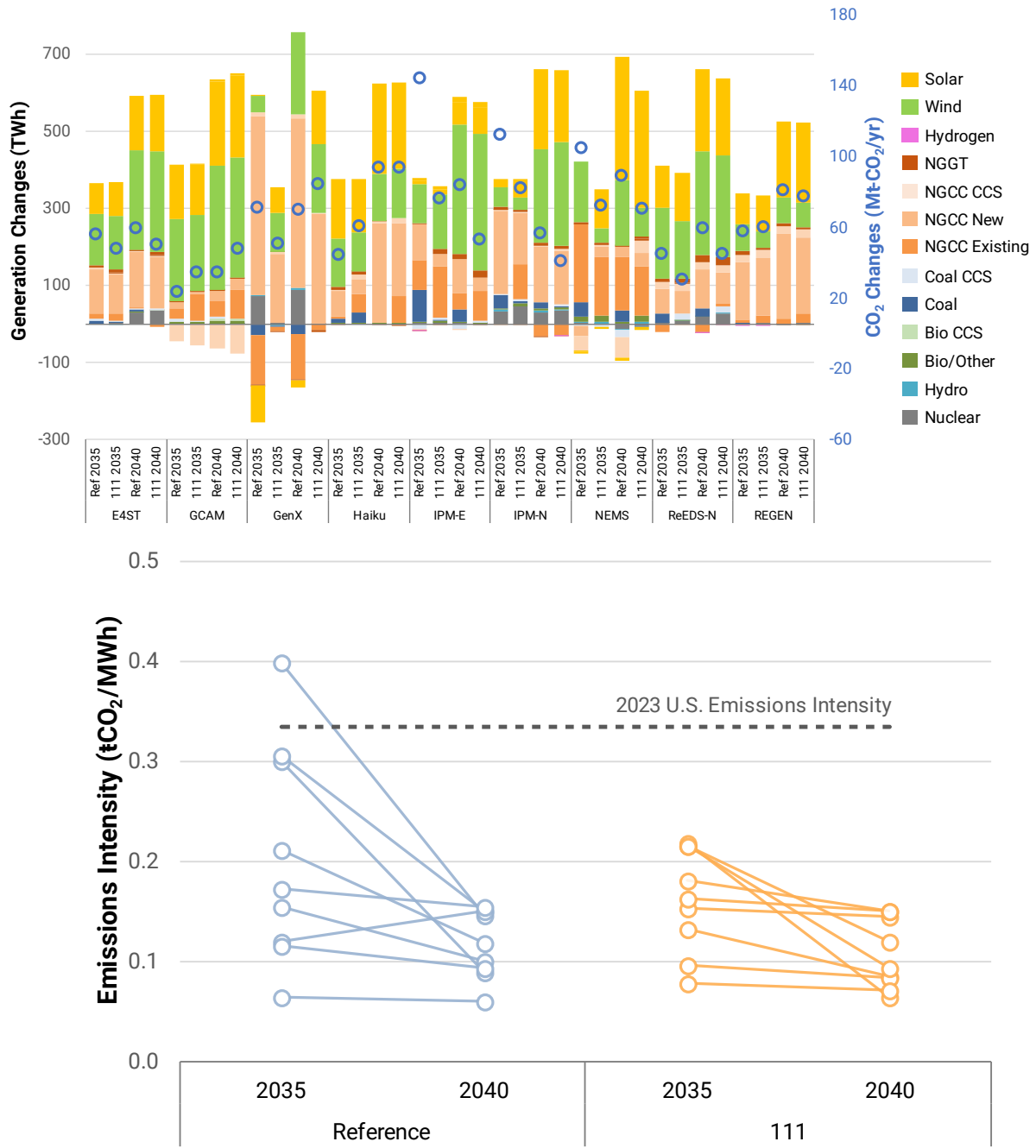

**Fig. S17. Incremental impacts of higher load growth on the generation mix and CO$_2$ emissions by model.** The top panel shows generation changes by scenario (reference and 111) and by time period (2035 and 2040). The bottom panel shows the marginal CO$_2$ emissions intensity of electricity generation under the higher demand scenarios and the 2023 U.S. intensity.

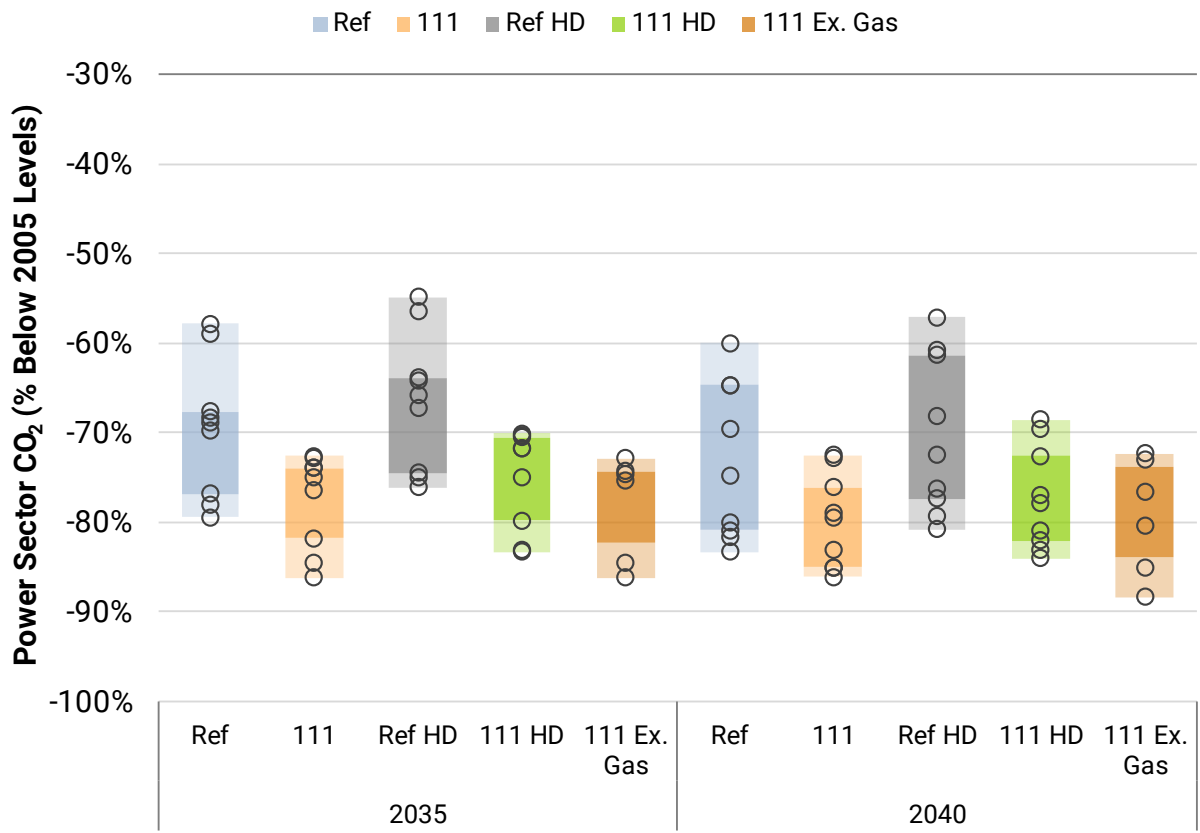

**Fig. S18. Cross-model comparison of U.S. power sector CO$_2$ emissions reductions by scenario relative to 2005 levels.** Circles show individual model results, while ranges show minimum and maximum values across models. "HD" refers to scenarios with higher electricity demand. "111 Ex. Gas" is the scenario that includes 111(d) rules for existing gas capacity. Ranges across models show the difference between the minimum and maximum values (lighter) and interquartile range (darker). Details on scenarios assumptions are provided in SM S3.

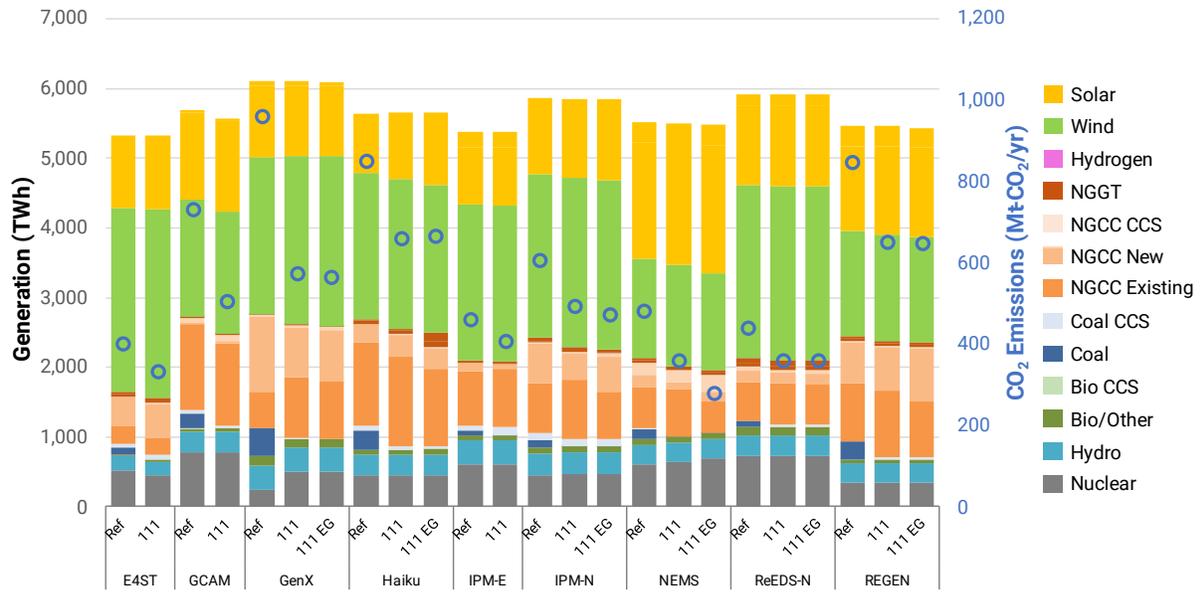

**Fig. S19. Cross-model comparison of generation in 2040 by model.** Modeled results show scenarios without and with EPA's 111 rules (Reference and 111, respectively) as well as a scenario that includes 111 rules with standards for existing gas power plants (see SM S3 for scenario details). CCS = carbon capture and sequestration; NGCC = natural gas combined cycle; NGGT = gas turbines (including all non-NGCC gas capacity).

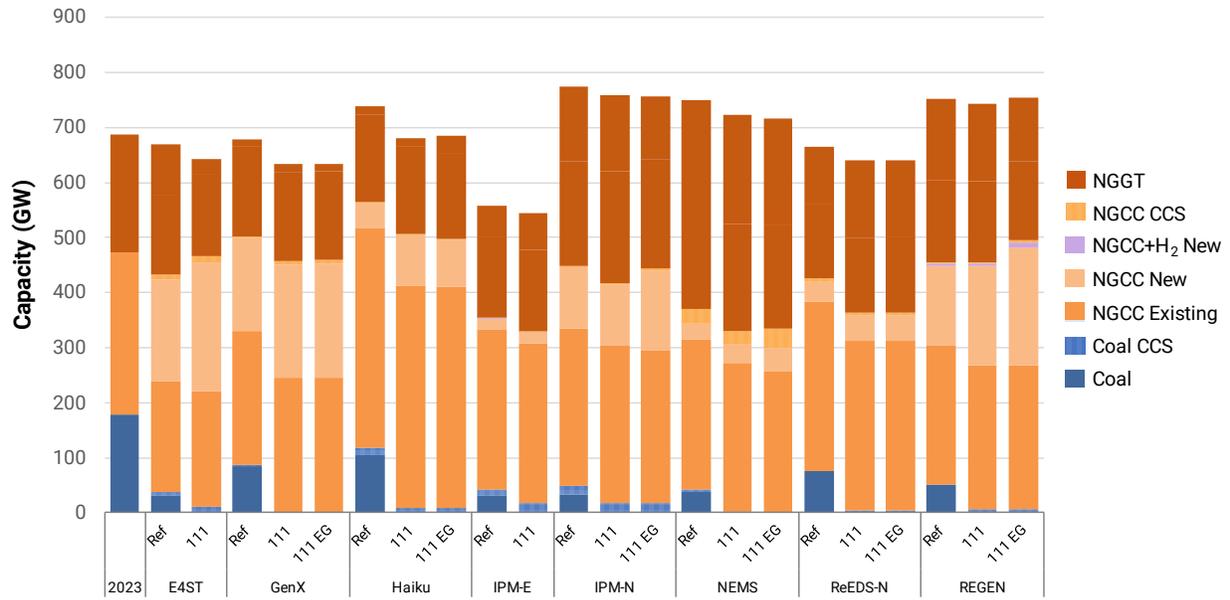

**Fig. S20. Cross-model comparison of fossil capacity in 2040 by model.** Modeled results show scenarios without and with EPA's 111 rules (Reference and 111, respectively) as well as a scenario that includes 111 rules with standards for existing gas power plants (see SM S3 for scenario details). CCS = carbon capture and sequestration; NGCC = natural gas combined cycle; NGGT = gas turbines (including all non-NGCC gas capacity).

**Table S1.**

**Participating models and policies represented in the reference scenario.** The Inflation Reduction Act (IRA) section is shown in parentheses.

| Category | Policy | E4ST-RFF | GCAM-CGS | GenX-ZERO | Haiku-RFF | IPM-EPA | IPM-NGO | NEMS-RHG | ReEDS-NREL | REGEN-EPRI |
|---|---|---|---|---|---|---|---|---|---|---|
| **IRA Provisions** | Production tax credit (PTC) extension (13101) | Included | Included | Included | Included | Included | Included | Included | Included | Included |
| | Investment tax credit (ITC) extension (13102) | Included | Included | Included | Included | Included | Included | Included | Included | Included |
| | Solar in low-income communities (13103/13702) | Included | Not Included | Not Included | Included | Included | Included | Not Included | Included | Included |
| | PTC for existing nuclear (13015) | Included | Included | Included | Included | Included | Included | Included | Included | Included |
| | New clean electricity PTC (45Y, 13701) and ITC (48E, 13702) | Included | Included | Included | Included | Included | Included | Included | Included | Included |
| | 45Q: Extension of credits for captured CO2 (13104) | Included | Included | Included | Included | Included | Included | Included | Included | Included |
| | 45V: Production credits for clean hydrogen (13204) | Included | Included | Not Included | Not Included | Included | Included | Included | Included | Not Included |
| **Other Policies** | Infrastructure Investment and Jobs Act (IIJA) | Included | Included | Included | Included | Included | Included | Included | Included | Included |
| | State Renewable Portfolio Standards | Included | Included | Included | Included | Included | Included | Included | Included | Included |
| | State Clean Electricity Standards | Included | Not Included | Included | Included | Included | Included | Not Included | Included | Included |
| | State Solar Mandates | Included | Not Included | Not Included | Included | Included | Included | Included | Included | Included |
| | State Offshore Wind Mandates | Included | Included | Included | Included | Included | Included | Included | Included | Included |
| | State Energy Storage Mandates | Not Included | Not Included | Included | Included | Included | Included | Included | Included | Included |
| | Regional Greenhouse Gas Initiative (RGGI) | Included | Included | Included | Included | Included | Included | Included | Included | Included |
| | State Carbon Caps (Power Sector or Economy) | Included | Included | Included | Included | Included | Included | Included | Included | Included |
| | State Nuclear Moratoria | Included | Not Included | Not Included | Not Included | Not Included | Not Included | Included | Included | Included |
| | Other EPA Regulations | Included | Not Included | Not Included | Included | Included | Included | Included | Included | Included |

**Table S2.**

**Representation of fossil fuel generation options and emissions control options across participating models.** The coverage is shown for existing and new coal and gas capacity.

| Fuel | Technology | E4ST-RFF | GCAM-CGS | GenX-ZERO | Haiku-RFF | IPM-EPA | IPM-NGO | NEMS-RHG | ReEDS-NREL | REGEN-EPRI |
|---|---|---|---|---|---|---|---|---|---|---|
| Coal | Existing coal | Included | Included | Included | Included | Included | Included | Included | Included | Included |
| | Existing coal: CCS retrofit | Included | Included | Included | Included | Included | Included | Included | Included | Included |
| | Existing coal: Gas cofiring | Included | Not Included | Included | Included | Included | Included | Not Included | Included | Included |
| | Existing coal: Gas conversion | Included | Not Included | Not Included | Not Included | Included | Included | Included | Not Included | Included |
| | Existing coal: Biomass cofiring | Not Included | Not Included | Not Included | Not Included | Included | Included | Included | Included | Included |
| | Existing coal: Biomass conversion | Not Included | Not Included | Not Included | Not Included | Included | Included | Included | Included | Included |
| | New coal: 90% CCS | Not Included | Included | Included | Not Included | Included | Included | Included | Included | Included |
| | New coal: >90% CCS | Not Included | Included | Not Included | Not Included | Included | Included | Not Included | Included | Included |
| Gas | Existing NGCC | Included | Included | Included | Included | Included | Included | Included | Included | Included |
| | Existing NGCC: CCS retrofit | Included | Included | Included | Included | Included | Included | Included | Included | Included |
| | Existing NGCC: Hydrogen cofiring | Not Included | Included | Included | Not Included | Included | Included | Not Included | Not Included | Included |
| | New NGCC | Included | Included | Included | Included | Included | Included | Included | Included | Included |
| | New NGCC: 90% CCS | Included | Included | Not Included | Included | Included | Included | Included | Included | Included |
| | New NGCC: >90% CCS | Not Included | Included | Included | Not Included | Not Included | Included | Included | Included | Included |
| | New NGCC: Hydrogen cofiring | Not Included | Included | Included | Not Included | Included | Included | Not Included | Not Included | Included |
| | Existing CT | Included | Included | Included | Included | Included | Included | Included | Included | Included |
| | Existing CT: Cofiring with low-emitting fuel | Not Included | Included | Included | Not Included | Included | Included | Not Included | Included | Included |
| | New CT | Included | Included | Included | Included | Included | Included | Included | Included | Included |
| | New CT: Cofiring with low-emitting fuel | Not Included | Not Included | Included | Not Included | Included | Included | Not Included | Included | Included |

**Table S3.**
**Key characteristics of models.** Temporal resolution is the number of intra-annual segments for power sector dispatch and investments. IPM-EPA results come from EPA's Regulatory Impact Analysis for the finalized 111 rules.

| Analysis Abbreviation | Model(s) | Analysis Institution | Model Type | Geographic Coverage | Spatial Resolution | Temporal Resolution | Documentation |
|---|---|---|---|---|---|---|---|
| **E4ST-RFF** | Engineering, Economic, and Environmental Electricity Simulation Tool | Resources for the Future | Electric sector | Contiguous U.S. and Canada | 6,000 transmission nodes | 16 days | [Link] |
| **GCAM-CGS** | Global Change Analysis Model | UMD-CGS | Energy systems | 50 U.S. states and D.C. | States | 4 segments | [Link] |
| **GenX-ZERO** | GenX | Princeton ZERO Lab | Electric sector | Contiguous U.S. | 26 regions | 4,368 segments | [Link] |
| **Haiku-RFF** | Haiku Power Sector Model | Resources for the Future | Electric sector | Contiguous U.S. | States | 24 segments | [Link] |
| **IPM-EPA** | Integrated Planning Model | EPA | Electric sector | Contiguous U.S. | 67 regions | 24 segments | [Link] |
| **IPM-NGO** | Integrated Planning Model | NRDC and CAELP | Electric sector | Contiguous U.S. | 67 regions | 24 segments | [Link] |
| **NEMS-RHG** | National Energy Modeling System | Rhodium Group | Energy systems | 50 U.S. states and D.C. | 25 regions | 9 segments | [Link] |
| **ReEDS-NREL** | Regional Energy Deployment System | NREL | Electric sector | Contiguous U.S. | 134 regions | 33 segments | [Link] |
| **REGEN-EPRI** | Regional Economy, Greenhouse Gas, and Energy | EPRI | Energy systems | Contiguous U.S. | 16 regions | 120 segments | [Link] |

**Table S4.**
**Technology-specific model constraints.** See Table S3 for a summary of participating models.

| Analysis Abbreviation | Wind and Solar | Transmission | Nuclear | CCS | Other Generation Options |
|---|---|---|---|---|---|
| **E4ST-RFF** | Site-specific resource quality (NREL) and transmission grid constraints | No endogenous transmission expansion, other than spur lines to connect new generation | No endogenous expansion | Transport and sequestration cost model from IPM, plus $35/t-$CO_2$ through 2035 to represent scaling costs | New gas capacity has 15% cost adder in places without historical capacity |
| **GCAM-CGS** | Bounds on regional resource quality (resource curves from ReEDS) | None | Constraint that only states that have built nuclear in past can expand | Constraint on near-term deployment for power; bounds on $CO_2$ storage | Hydro is fixed at historical levels |
| **GenX-ZERO** | Bounds on regional resource quality | None | Nuclear plants cannot retire before 2035 | $CO_2$ storage and transport costs vary by region | None |
| **Haiku-RFF** | None | Constraint on interstate transmission capacity growth (upper bound of 0.5% per year) | Only planned nuclear additions are allowed | Upper bound on $CO_2$ storage of 100 million short tons in 2030, doubling every five years thereafter | Capacity of each fuel type is limited to historical maximum growth increasing at 7% annually |
| **IPM-EPA** | Bounds on regional resource quality and potential based on NREL; planned units through 2028 | Endogenous transmission builds starting 2030 between contiguous regions | Includes exogenous projections of plant capacity factors; while planned retirements are included, endogenous economic retirements are not allowed | CCS available starting in 2030; includes sequestration cost curves and capacity by location; transport links between $CO_2$ source and sink capped at 750 miles | None |
| **IPM-NGO** | Bounds on regional resource quality and potential based on NREL; includes short-term capital adder ($/kW) for levels of projected annual deployment significantly above historical record through 2035 | Endogenous transmission builds starting 2028 between contiguous regions | Regulated nuclear plants can only be endogenously retired after the age of 60; endogenous plant capacity factors must be at least 80% | CCS builds in 2028 and 2030 are limited to plants that have already begun FEED study or announced plans to install CCS; includes sequestration cost curves and capacity by location; transport links between $CO_2$ source and sink capped at 750 miles | None |
| **NEMS-RHG** | Bounds on regional resource quality; lower bounds for planned additions (EIA-860) | Interregional transmission constraints are derived from Form EIA-411 | Constraint on near-term deployment and state-level policies | Constraint on near-term deployment for power | Lower bounds to reflect under-construction capacity (EIA-860) |
| **ReEDS-NREL** | Bounds on regional resource quality; lower bounds for planned additions through 2024; growth penalty based on | Near-term announced additions; before 2028, endogenous expansion is limited to historical | State-level policies represented; nuclear plants cannot retire until 2032 | Retrofits are not allowed until 2028 or later | None |
| | previous year through 2034[18] | maximum build rate (1.4 TW-mi/yr) | | | |
| **REGEN-EPRI** | Bounds on regional resource quality; lower bounds for planned additions (EIA-860) | Limits on growth of inter-regional capacity based on earlier capacity | Constraint on near-term deployment and state-level policies; nuclear plants cannot retire until 2032[19] | Constraint on near-term deployment for power; bounds on $CO_2$ storage | Lower bounds to reflect under-construction capacity (EIA-860) |

---

[18] This growth penalty in ReEDS applies to all technologies but is only binding for wind and solar.
[19] REGEN assumes that existing nuclear capacity does not retire before 60 years when it is majority owned by a utility with a net-zero emissions target.

**Table S5.**
**Representation of non-economic factors.** See Table S3 for a summary of participating models.

| Analysis Abbreviation | Siting and Land Use Restrictions | Federal Lands Representation | Interconnection Process Representation | Supply Chains and Workforce | Public Opinion |
|---|---|---|---|---|---|
| **E4ST-RFF** | NREL site data screened based on population density thresholds for wind/solar | Restrictions based on NREL | None | Ex-post analysis of scenario-specific impacts | None |
| **GCAM-CGS** | Applying site-level exclusions for land cover, urban areas, slope, and restricted areas, which gives regional wind and solar resource capacity constraints by class | None | Static interconnection costs are included but process is not explicitly modeled | None | None |
| **GenX-ZERO** | Utility-scale solar PV, onshore wind, and offshore wind candidate project areas are identified | Excluded federally protected land | Interconnection costs varying by location are included but process is not explicitly modeled | None | None |
| **Haiku-RFF** | None | None | None | None | None |
| **IPM-EPA** | Site-level exclusions as represented in NREL resource data for wind and solar | Restricted areas on certain types of land as represented in NREL resource data for wind and solar | Interconnection costs are included but process is not explicitly modeled | Regional multipliers account for differences in labor cost | None |
| **IPM-NGO** | Site-level exclusions as represented in NREL resource data for wind and solar | Restricted areas on certain types of land as represented in NREL resource data for wind and solar | Interconnection costs are included but process is not explicitly modeled | Regional multipliers account for differences in labor cost | None |
| **NEMS-RHG** | Applying [site-level exclusions](#) for land cover, urban areas, slope, and restricted areas, which gives regional wind and solar resource capacity constraints by class | Restricted areas on certain types of federal lands represented for wind | None | None | None |
| **ReEDS-NREL** | Applying site-level exclusion for land cover, urban areas, slope, ridgelines, habitat protected areas, restricted areas, and more from [NREL's reV model](#) | Costs, construction timelines, etc. are same on federal and private lands, though forthcoming update will alter land availability | Interconnection costs are included but process is not explicitly modeled | Not explicitly represented, though growth penalties are included to represent scalability challenges, and regional multipliers account for differences in labor cost | None |

| **REGEN-EPRI** | Applying [site-level exclusions](#) for land cover, urban areas, slope, and restricted areas, which gives regional wind and solar resource capacity constraints by class | Restricted areas on certain types of federal lands represented | Interconnection costs are included but process is not explicitly modeled | Ex-post analysis of scenario-specific impacts; regional multipliers account for differences in labor cost | None |

**Table S6.**
**Summary of key indicators for 111 and reference scenarios across models in 2035.** Values are visualized in Fig. 2. Indicators are electric sector $CO_2$ reductions (% from 2005 levels), generation share from coal without CCS, generation share from natural gas without CCS, CCS-equipped generation share, and wholesale electricity price changes relative to the reference scenario.

| Metric | Units | E4ST-RFF | GCAM-CGS | GenX-ZERO | Haiku-RFF | IPM-EPA | IPM-NGO | NEMS-RHG | ReEDS-NREL | REGEN-EPRI |
|---|---|---|---|---|---|---|---|---|---|---|
| **Ref: Electric Sector $CO_2$ Reduction** | % from 2005 | 77% | 68% | 58% | 68% | 70% | 69% | 79% | 78% | 59% |
| **111: Electric Sector $CO_2$ Reduction** | % from 2005 | 82% | 76% | 74% | 73% | 75% | 74% | 86% | 85% | 73% |
| **Ref: Unabated Coal Generation Share** | % | 3.5% | 4.0% | 8.5% | 4.5% | 3.2% | 2.8% | 4.0% | 3.4% | 8.3% |
| **111: Unabated Coal Generation Share** | % | 0.3% | 0.5% | 0.9% | 2.2% | 0.1% | 0.0% | 0.2% | 0.0% | 0.0% |
| **Ref: Unabated Gas Generation Share** | % | 19% | 25% | 28% | 28% | 27% | 31% | 14% | 17% | 27% |
| **111: Unabated Gas Generation Share** | % | 21% | 25% | 29% | 28% | 28% | 31% | 15% | 17% | 31% |
| **Ref: CCS-Equipped Generation Share** | % | 1.4% | 1.8% | 0.6% | 2.0% | 1.6% | 2.4% | 3.4% | 0.8% | 0.3% |
| **111: CCS-Equipped Generation Share** | % | 2.1% | 1.9% | 0.7% | 1.5% | 2.8% | 2.8% | 4.3% | 1.5% | 1.2% |
| **111: Electricity Price Change from Reference** | % from Ref | 0.8% | | 1.5% | 0.1% | 6.6% | 6.6% | 0.1% | 7.1% | 0.8% |